\documentclass[letter,11pt]{article}
\pdfoutput=1
\usepackage[margin=2cm]{geometry}
\usepackage[sumlimits,intlimits]{amsmath}
\usepackage{amsfonts,amssymb}
\usepackage{mathrsfs}
\usepackage{amsmath}
\usepackage[T1]{fontenc}
\usepackage{capt-of}
\usepackage{ucs}
\usepackage{graphics}
\usepackage[dvips]{graphicx}
\usepackage{textcomp}
\DeclareGraphicsExtensions{.pdf,.png,.eps}
\usepackage{lineno}
\usepackage[square,sort&compress,comma]{natbib}
\setcitestyle{numbers} 

\usepackage{lineno}
\newenvironment{mylisting}
{\begin{list}{}{\setlength{\leftmargin}{1em}}\item\scriptsize\bfseries\linenumbers}
{\end{list}}

\begin{document}
\makeatletter
\@addtoreset{equation}{section}
\makeatother
\renewcommand{\theequation}{\thesection.\arabic{equation}}


\vspace{.5cm}
\begin{center}
\Large{\bf  Slow-Roll Inflation in Non-geometric Flux Compactification}\\

\vspace{1cm}

\large
Cesar Damian\footnote{cesaredas@fisica.ugto.mx}, Luis R. D\'iaz-Barr\'on\footnote{luisreydb@fisica.ugto.mx}, Oscar Loaiza-Brito\footnote{oloaiza@fisica.ugto.mx} and  M. Sabido\footnote{msabido@fisica.ugto.mx} \\[4mm] 
{\small\em Departamento de F\'isica, Universidad de Guanajuato,}\\
{\small\em  C.P. 37150, Leon, Guanajuato, Mexico.}\\[4mm]
\vspace*{2cm}
\small{\bf Abstract} \\
\end{center}

\begin{center} 
\begin{minipage}[h]{14.0cm} {By implementing a genetic algorithm we search for stable vacua in Type IIB non-geometric flux compactification on an isotropic torus with orientifold 3-planes.  We find that the number of stable dS and AdS vacua are of the same order. Moreover we find that in all dS vacua the multi-field slow-roll inflationary conditions are fulfilled. Specifically  we observe that inflation is driven by the axio-dilaton and the K\"ahler moduli. We also comment on the existence of one stable dS vacuum in the presence of exotic orientifolds. }

\end{minipage} 
\end{center}

\bigskip

\bigskip
 
\bigskip
 


\newpage

\section{Introduction}
Recently, there has been a huge interest in the search for classical de Sitter (dS) vacua within the context of superstring compactification. In the last few years, some constraints have been imposed by estimating the contributions to the effective scalar potential from each  of the components that play a role in the compactification process.  For instance,  there is a no-go theorem, indicating that the existence of dS vacua in compactifications threaded with standard NS-NS and R-R fluxes is incompatible with inflation\cite{Hertzberg:2007wc}. Recently, in the context of these standar type IIB compactifications, it has been shown the existence of classical dS vacua in specific and suitable D-brane configurations with orientifold planes \cite{Cicoli:2012vw, MartinezPedrera:2012rs, Louis:2012nb}.  On the other hand, further studies show that by considering a bigger set of allowed fluxes and more general structures for the internal geometry, it is possible to find some stable dS vacua \cite{Chen:2011ac,  Shiu:2011zt, Danielsson:2011au, Danielsson:2009ff, Saltman:2004jh, Haque:2008jz, Cicoli:2012fh,  Danielsson:2010bc,  Covi:2008ea, Silverstein:2007ac,  Dong:2010in, Dong:2010pm,  Kachru:2003aw,  Kallosh:2010xz} although their compatibility with inflation has not been studied in detail. In this work we present  some stable  dS vacua consistent with inflation.\\

There are some essential ingredients  string compactifications should contain to enhance the chances of finding stable dS vacua:
 a negative curved internal manifold  and a small number of moduli fields \cite{Douglas:2010rt, Marsh:2011aa,  Baumann:2009ni}.\\

Here, we  take the approach consisting in computing the scalar potential from a superpotential with all the above features.  Specifically, a compactification on a negatively curved manifold with a superpotential $\mathscr{W}$ that depends  at tree level on a small set of moduli is achieved by considering a type II string compactification on a six-dimensional isotropic torus  in the presence of non-geometric fluxes  \cite{Wecht:2007wu, Shelton:2006fd, Shelton:2005cf, Flournoy:2004vn}.\\

These type of compactifications have become a fruitful region in the string landscape  to look for stable dS vacua \cite{Aldazabal:2006up, Camara:2005dc}. For instance, by using an algebraic geometry approach  \cite{Dibitetto:2012ia, Guarino:2008ik,  Font:2008vd, deCarlos:2009qm, Conlon:2010ji, deCarlos:2009fq},  it has been shown that in the presence of non-geometric fluxes in Type IIB compactifications on $T^6/\mathbf{Z}_2\times\mathbf{Z}_2$, there are stable dS vacua which are continuously connected to Minkowski non-supersymmetric vacua. Furthermore, recently the distribution of dS vacua and the presence of tachyon directions were studied in \cite{Danielsson:2012et, Danielsson:2012by} and a classification of stable vacua was shown in \cite{Blaaback:2013ht} for a compactification on an anisotropic torus.\\

As important as the search of dS vacua is, it is tempting to analyze these models in the context of inflation. Are there suitable conditions for slow-roll inflation in these models?
 Huge advances have been made in this field by considering general string compactifications \cite{deCarlos:1993jw, Banks:1995dp, Cicoli:2007xp, Conlon:2007gk, Balasubramanian:2005zx, Cicoli:2010yj, Cicoli:2011ct, Baumann:2009ni,  Cicoli:2011zz, Kallosh:2007ig, Borghese:2011en}.  One of the first models was  studied in \cite{Hertzberg:2007wc},  where it was proved that in the context of a standard vanilla compactification, the slow-roll cosmological parameters were much bigger than 1, excluding the possibility of inflation. Other models considered the presence of extra features,   as more fluxes and extra stringy objects (orientifolds) yielding the construction of inflationary scenarios. However, recently it was shown that single field inflation is forbidden \cite{Borghese:2012yu} in the small-field regime.\\

Explicitly computing the slow-roll parameters in specific models (as the non-geometric tori) requires numerical calculations. It is the purpose of this work to implement a genetic algorithm to search for stable dS vacua\footnote{Previous advances towards this goal were shown in \cite{Damian:2012yp}.} and compute the corresponding slow-roll parameters,  in order to establish if inflation is viable within these models.\\

We find  several stable vacua with positive and negative values at their minimum; interestingly we observe that the number of {\it stable} dS and Anti-de Sitter (AdS) vacua are of the same order. By computing the corresponding masses, we see that there is a hierarchy,  pointing out the possibility to have  multi-field inflation driven by a subset of the moduli. Specifically,  we find that the inflaton is driven by the axio-dilaton and the K\"ahler parameter,  with a direction approximately orthogonal to the s-goldstino.  We observe that all stable dS vacua contain conditions for small multi-field driven slow-roll inflation. In particular we find that inflation is mainly driven by the internal volume and the real part of the dilaton field.\\

We also find that supersymmetry is broken through the complex structure at a trans-Planckian scale (with a high value of the gravitino mass for a realistic model). Finally we show that a single field scenario for inflation is not allowed,  in agreement with the results shown in \cite{Borghese:2012yu}. We also comment on the existence of stable dS vacua in the presence of exotic orientifolds, an issue that has been so far avoided in literature.  We believe that by considering exotic orientifolds it is possible to increase  the moduli field space,  where dS vacua can be found.\\

It is important to remark that still it is not clear whether these models are constructed from a ten-dimensional scenario. The topic is still under study and great efforts have been made over the past few years \cite{Andriot:2010ju, Andriot:2011uh}. The outline of our paper is as follows:  in Section 2 we  fix our notation and study which moduli break SUSY while fulfilling the tadpole constraints and Bianchi identities. Section 3 is devoted to the implementation of the genetic algorithm to look for stable dS and AdS vacua. In section 4 we compute the slow-roll parameters and  we show the possibility of inflationary trajectories for small fields. Also, we show that stable dS vacua are present in a background with exotic orientifolds. In section 5 we present our final comments and conclusions. \\


{\bf Note added for Version 3 in ArXiv}:  After being published we detect a mistake concerning Eq. (2.7), which was not complete. In this version we update the equation in terms of the complex superpotential $\mathscr{W}$ and,  in Appendix A,  we have added the explicit form of the scalar potential for the cases presented in Table 2 and for the general case in which all fluxes are considered.  Similarly, some  typos in Table 1 and Table 2 were corrected related to the nomenclature we used in computing the scalar potential. Finally we want to remark that {\it the results and conclusions stated in our previous version are unchanged}.


\section{Classical dS vacua}
We  look for classical stable (A)dS vacua by considering  a Type IIB compactification on a six-dimensional isotropic torus $T^6$ which is  constructed by three identical copies of two-dimensional torus $T^2$ threaded with RR, NS-NS and non-geometric fluxes, under a $\mathbf{Z_2}$ symmetry given by the presence of  O3-planes  \cite{Shelton:2005cf}.   The effective $N=1, D=4$ theory contains three complex
moduli, namely the axio-dilaton $S$, a complex structure modulus  $\tau$ and a complexified K\"ahler modulus $U$.
The contribution of the non-geometric fluxes to the superpotential is obtained by adding a term to the well known Gukov-Vafa-Witten \cite{Aldazabal:2006up}
\begin{equation}
\mathscr{W}(\tau, S, U)= \int_{T^6/\mathbf{Z}_2} \left({F_3-S H_3 - Q \cdot U}\right)\wedge\Omega,
\end{equation}
where $Q\cdot U=Q^{mn}_{[q}U_{mnrs]}dx^q\wedge dx^r\wedge dx^s$ is a 3-form and $\Omega$ is the standard holomorphic 3-form. \\
 
 In such scenario the tree-level  superpotential $\mathscr{W}$ depends on all moduli and after integration it is given by
\begin{equation}
\mathscr{W}(\Phi_i)=P_1(\Phi_1)+\sum_{i=2}^3 \Phi_i P_i(\Phi_1),
\end{equation}
where we have  fixed the notation as follows: $\Phi_i=\phi_i+i\psi_i=\{\tau, S, U\}$ are the complex moduli fields and $P(\Phi_1)$ is a polynomial of cubic order on  $\Phi_1$ with real coefficients $A_{ij}$,
\begin{equation}
P_i(\Phi_1)=\sum_j A_{ij}\Phi_1^j.
\label{poly}
\end{equation}
The real coefficients $A_{ij}$ in (\ref{poly})  are actually the integrated fluxes on the appropriate cycles of $T^6$ (see Appendix \ref{ap:notation} for notation) given by
\begin{equation}
A_{ij}=\left(
\begin{array}{cccc}
a_{00}&-\sum_k a_{01}^k&\sum_k a^k_{02}&-a_{03}\\
&&&\\
a_{10}&-\sum_ka^k_{11}&\sum_ka^k_{12}&-a_{13}\\
&&&\\
a_{20}&\sum_ka_{21}^k&-\sum_k a_{22}^k&-a_{23}
\end{array}\right).
\label{polymatrix}
\end{equation}
The coefficients $\{a_{0i}\}$ correspond to the integrated  RR-fluxes $F_3$ , $\{a_{1i}\}$ to NS-NS fluxes $H_3$ and $\{a_{2i}\}$ to non-geometric fluxes $Q$ all over the 3-cycles on $T^6$.  Notice that for the isotropic case we are considering, 
\begin{equation}
\sum_{k=1}^3 a^k_{ji}=3a_{ji}, \qquad \text{for $j=0,1$}.
\end{equation}
We also consider the usual K\"ahler potential $\mathscr{K}$ of the form
\begin{equation}
\mathscr{K}=-\sum_{i=1}A_i~log(2\psi_i),
\end{equation}
with positive $A_i=\{3,1,3\}$.
Hence, if SUSY is broken by one or several moduli, the scalar potential reads
\begin{equation}
e^{-\mathscr{K}}\mathscr{V}=\sum_{i}\left(4\frac{\psi_i^2}{A_i}|\partial_i \mathscr{W}|^2+4\psi_i~Im~(\bar{\mathscr{W}}\partial_{\bar{i}}\mathscr{W}) +(A_i-3)|\mathscr{W}|^2\right),
\label{scalarp}
\end{equation}
where the sum runs over the moduli that break SUSY. \\

We can  see that it is possible to obtain positive values for $\mathscr{V}$ at its minimum.  However, the existence of dS classical solutions depends on the presence of the non-geometric fluxes $a_{2j}$. If they are turned off not only it is impossible to stabilize the K\"ahler modulus but also there is no  possibility to find classical stable dS solutions since
\begin{equation}
\frac{\partial\mathscr{V}}{\partial\psi_3}\sim \frac{1}{\psi_3^4},
\end{equation}
which vanishes as $\psi_3\rightarrow \infty$ together with the whole potential, rendering the space-time to be flat.  A similar situation holds in the absence of RR and NS-NS fluxes since in this case the scalar potential also contains a run-away direction, 
\begin{equation}\frac{\partial\mathscr{V}}{\partial\psi_2}\sim \frac{1}{\psi_2^2},
\end{equation}
that vanishes as  $\psi_2\rightarrow\infty$,  taking our model far away from the perturbative regime. Therefore, it is necessary  to at least, turn on NS-NS, R-R and non-geometric fluxes, for the model to avoid the above run-away directions.\\

The required conditions for dS or AdS vacua depends on which moduli breaks supersymmetry and how the vanishing of $F$-terms is compatible with Bianchi identities and tadpole constraints.

\subsection{Constraints}

As it is known, the presence of fluxes contributes with an amount of internal energy which must be cancelled by the addition of a negative tensioned object. These requirements lead to a set of constraints the fluxes must satisfy, as the R-R tadpole constraint $\frac{1}{2}\int H_3\wedge F_3 + \tilde{N}=16$ and its T-dual counterpart $F_3\cdot Q=0$ which after integration are given by
\begin{equation}
\sum_k\sum_{i=0}^3 (-1)^{i}a^k_{0i}a^k_{j(3-i)}=2(16-\widetilde{N})\delta_{1j}, \qquad \text{for $j=1,2$,}
\end{equation}
where the sum over $k$ stands for the corresponding  elements in the matrix $a_{ij}$, and  $\widetilde{N}$ is the number of exotic orientifold planes. An odd number of them are required at the fixed points for every odd flux supported in a cycle \cite{Frey:2002hf}; the extra constraints are encoded in the Bianchi identities. With the purpose to rewrite them, let us consider the vectors $\vec{A}^M_J:=\vec{A}_{j\pm,k}^{m\pm}$  given by
\begin{eqnarray}
\vec{A}_{j, k\pm}^{m\pm}&=&(a^{m}_{j(k)}, a^{m\pm1}_{j(k\pm1)}, a^{m\pm2}_{j(k\pm 1)}, a^{m\pm 3}_{j(k\pm 2)}),\\
\end{eqnarray}
where $J=(j\pm,k)$ and $M=(m\pm)$. 
The  Bianchi identity  $Q\cdot H_3=0$ decomposes as 
\begin{eqnarray}
\vec{A}_J^M\cdot \vec{A}_{J'}^{M'}=(A_J^M)_\delta(A_{J'}^{M'})_\lambda\eta^{\delta\lambda}=0,
\end{eqnarray}
for $j\neq j'$, $diag(\eta)=\{-1,1,1,1\}$  and for the combinations   $M=(0+), J=\{ (2,0+), (1,3-)\}$ and $ M=(3-), J=\{(1, 2-), (2,1+)\}$, while  $Q\cdot Q=0$ decomposes as,
\begin{equation}
\vec{C}_i\cdot \vec{C}_j=C_{im}C_{jm}(-1)^m=0, 
\end{equation}
for $(i,j)=\{(1,2),(1,3),(3,4),(2,4)$, with
\begin{eqnarray}
\vec{C}_1= (a_{23},a^3_{22}, a^1_{22},a^2_{21}),\nonumber\\
\vec{C}_2=(a^1_{21},a^3_{22},a^2_{22},a_{23}),\nonumber\\
\vec{C}_3=(a_{20},a^2_{21}, a^3_{21}, a^1_{22}),\nonumber\\
\vec{C}_4=(a^2_{22}, a^1_{21}, a^3_{21},a_{20}).
\end{eqnarray}

A formal derivation of the above constraints is found in \cite{Shelton:2005cf}.\\

\subsection{SUSY breakdown} 
Since we are not considering the presence of D-branes, there are no D-terms in the potential, and 
 the breakdown of SUSY   is achieved only through F-terms. Our model depends on three complex moduli $\Phi_i$, implying that  there are  7 ways to break SUSY\footnote{A complete solution of the 
supersymmetric cases is presented in \cite{Shelton:2006fd} with AdS vacua as expected.}.  It is therefore necessary to analyze all of them and check if they are compatible with the Tadpole and Bianchi identity constraints.\\

\noindent
A. {\it Breaking SUSY through a single modulus}.

\begin{enumerate}
\item
 Consider that SUSY is broken only through the complex structure $\Phi_1$, i.e., $D_{2,3}\mathscr{W}=0$ and $D_1\mathscr{W}\neq 0$. The SUSY equations leads to the condition  $Im(P_2\bar{P}_3)=0$, which in principle can be  compatible with the Tadpole and Bianchi identities, establishing a way to 
stabilize $\Phi_1$. The solution of the SUSY equations of motion for $\Phi_{2,3}$ provides a stabilization for $\Phi_2$ but leaves unfixed  the K\"ahler modulus $\Phi_3$. Its stabilization
 depends on the value of  the scalar potential acquires at its minimum. We shall explore this possibility in more detail in the next section. 
  \item
Breaking SUSY only  through the K\"ahler modulus $\Phi_3$ implies that the SUSY conditions $D_{1,2}\mathscr{W}=0$,
are fulfilled by any value of the moduli, even away from the minima. These two equations give the conditions
\begin{eqnarray}
\psi_1\partial_1P_i&=&0, \qquad  \text{for i=1,3},\\
i\psi_1\Phi_2\partial_1P_2-3\psi_2P_2&=&0.
\end{eqnarray}
Since we are assuming non-vanishing real moduli ($\phi_i$, $\psi_i$), the  second condition implies $P_2=0$ (which in turn says that we have turned off all NS-NS fluxes. This gives  a superpotential without a dependence on the axio-dilaton $\Phi_2$ and together with the first condition, it constraints the coefficients for $P_{1,2,3}$ to be related by $a_{0i}=\lambda_1 a_{1j}=\lambda_2 a_{2j}$ for each $j$ and for any integers $\lambda_{1,2}$. Then we obtain  a violation of the Tadpole conditions. Since such cases lead to non-physical conditions, we  exclude them from our analysis.
\item
Breaking SUSY through the dilaton modulus $\Phi_2$ leads to similar equations as the previous case and it will be also excluded from our analysis.
\end{enumerate}

\noindent
{\it Breaking SUSY through two moduli.} 
\begin{enumerate}
\item
Consider breaking SUSY through $\Phi_2$ and $\Phi_3$. Solutions to the equation $D_1\mathscr{W}=0$ imply that $P_1 ({\Phi_1})=P_2 ({\Phi_1})=P_3 ({\Phi_1})$. The 
fluxes associated with similar power-terms in $\Phi_1$ must have the same value. However this condition leads to
$a_{0j}=a_{1j}=\sum_ka^k_{2j}$ for all $j$,  and  violates the tadpole condition.
\item
Breaking SUSY through $\Phi_1$ and $\Phi_2$ or through $\Phi_1$ and $\Phi_3$ leads to $P_i=0$ for all $i$ which also violates the tadpole condition. 
\end{enumerate}
  All these three cases are also excluded from our analysis.\\
  
\noindent{\it Breaking SUSY through all the moduli} implies that all of them must be stabilized dynamically through
reaching the minima of the scalar potential. The values of the corresponding fluxes must satisfy the Bianchi identities
 and the moduli must acquire positive values. Since there are no constraints to consider, this case must be taken into account in our search for stable dS and anti-De-Sitter  vacua.\\

In summary, we have found that there are only two ways to obtain stable vacua,  compatible with Tadpole constraints, Bianchi identities and with the possibility to have a positive value at the minimum:  breaking SUSY through the complex structure modulus $\Phi_1$ or through all three complex moduli $\Phi_i$. It is important to remark that a generic flux configuration is not {\it ad hoc} compatible with these constraints. Therefore, we are interested in finding some specific  flux configurations which fulfill such constraints.

\subsection{Inflation in non-geometric compactifications}

As we have seen, the existence of (stable) vacua with small positive or negative energy values depends on the way SUSY is broken.  However, as shown by the expression  for the scalar potential in (\ref{scalarp}), it also depends on the flux configuration we consider and on the vacuum expectation values for the moduli. \\

Assuming the existence of such vacua (a task we face in the next section) it is of high interest to look for suitable conditions for the presence of inflation.  A rigorous proof for that   is quite complicated. However,  we can infer some features about the existence of such conditions.\\

Consider for instance a superpotential $\mathscr{W}=\mathscr{W}(\Phi_i)$ of the form
\begin{equation}
\mathscr{W}= P\sum_{i=1}^3Q_i(\Phi_i),
\end{equation}
for a constant complex $P$ and for $Q_i$ being a polynomial of order $n$ in $\Phi_i$ for all $i$. The corresponding scalar potential can be written as
\begin{equation}
e^{-\mathscr{K}}\mathscr{V}= \sum_i\left(\frac{4\psi^2}{A_1}|P|^2\right)+4W^2.
\end{equation}
It is clear that  the mass matrix is diagonal and the possibility to have all moduli masses at the same order is a viable option.  Therefore,  the possibility to identify  a candidate for the inflaton within a subset of the moduli fields is  excluded. Based on this argument, we think that a superpotential which does not look symmetric by interchanging the moduli fields $\Phi_i$ (modulo the flux values $a_{ij}$) indicates the possibility of having inflation driven by a subset of the moduli .  For instance, in KKLT-type models, it is customary assumed that dilaton and complex structure moduli fields are stabilized previous to the K\"ahler moduli, which acquire a vev by the presence of a non-perturbative term depending on them. In these models the inflaton is identified with the K\"ahler and the axio-dilaton moduli or rather a linear combination of them \cite{Kachru:2003aw}.\\

In our case, we have two (non-formally derived) reasons to believe that it is possible to identify a candidate for the inflaton with a single modulus complex field  or with two complex fields . First, the superpotential $\mathscr{W}$ looks symmetric on $\Phi_2$ and $\Phi_3$ but not on $\Phi_1$, therefore we expect the inflaton to be identified with the complex structure modulus $\Phi_1$ or with a combination of $\Phi_2$ and $\Phi_3$.  Second, we can show that in a generic scenario and with the expected conditions, the possibility that all masses are of the same order is very restricted, and in consequence,  there is a big possibility that the model has a hierarchy on the masses.
This points out the possibility to identify a candidate for the inflaton within a subset of the moduli fields. See  Appendix \ref{ap:mass} for details.\\

In the next section we shall concentrate on two issues: 1) searching for stable classical dS and AdS vacua on which supersymmetry is broken and 2) studying  the presence of suitable conditions for slow-roll inflation in the large and small field scenario. \\

\section{Searching for stable (Anti) De Sitter  vacua: the genetic algorithm}
Let us start by looking for stable dS and Anti-de Sitter classical vacua. Notice that the existence of a minimum with all moduli stabilized depends  on the values of the fluxes we are turning on. Looking for such vacua is a daunting task if one proceeds analytically. However it is possible to employ some computational algorithms which allows to find the existence of those vacua.\\

We apply a genetic algorithm\footnote{ A genetic algorithm is a probabilistic optimization method to search global minima that mimic the metaphor of natural 
biological evolution. The computational procedure operates on a set of initial values for the moduli (population) 
which gives a potential solution by applying the principle of survival of the fittest to produce successively better 
approximations to a minima of the scalar potential. At each generation of the numerical procedure, a new set of 
moduli is created by the process of selecting individuals according to their level of fitness in the moduli space
 and reproducing them using feasible adaptation. This process leads to the evolution of moduli that are better 
suited to the minimum value of the scalar potential than the set of moduli from which they were created, so the 
algorithm successively improves the solution. In Appendix C we show the code used in this calculation.}  with a convergence
criteria of $10^{-20}$. The numerical procedure is as follows. First
a set of integer fluxes are randomly generated satisfying the Tadpole and Bianchi constraints. Then the genetic algorithm is employed in order to find
the minima of the scalar potential.\\

Searching for minima requires to
 fix a bound for all the fluxes  \cite{Shelton:2005cf} ,
which then acts as a natural cutoff, leaving  the shape of the vacua distribution
unaltered as the bound increases. However, a genetic algorithm does not allow to find metastable vacua since it searches for the lowest value of the potential  in a neighborhood of a selected point. If another point is found in which the value of the scalar potential is lower than the previous one, the algorithm selects this point as the base  to look for new minima. Once the searching fails to find a lesser value, the algorithm stops. Also, if a minima is found, the algorithm will not continue looking for other minima for the same potential. Therefore, we cannot be certain that a minima is metastable or global.\\

In the next sections we shall employ the genetic algorithm to look for stable minima for  the two cases derived in the previous section. For completeness,  the algorithm code, with supersymmetry broken through all moduli,  is presented in Appendix \ref{ap:code}.

\subsection{Case 1: SUSY breakdown through the complex structure}

\begin{figure}
\begin{centering}
\includegraphics[width=7cm]{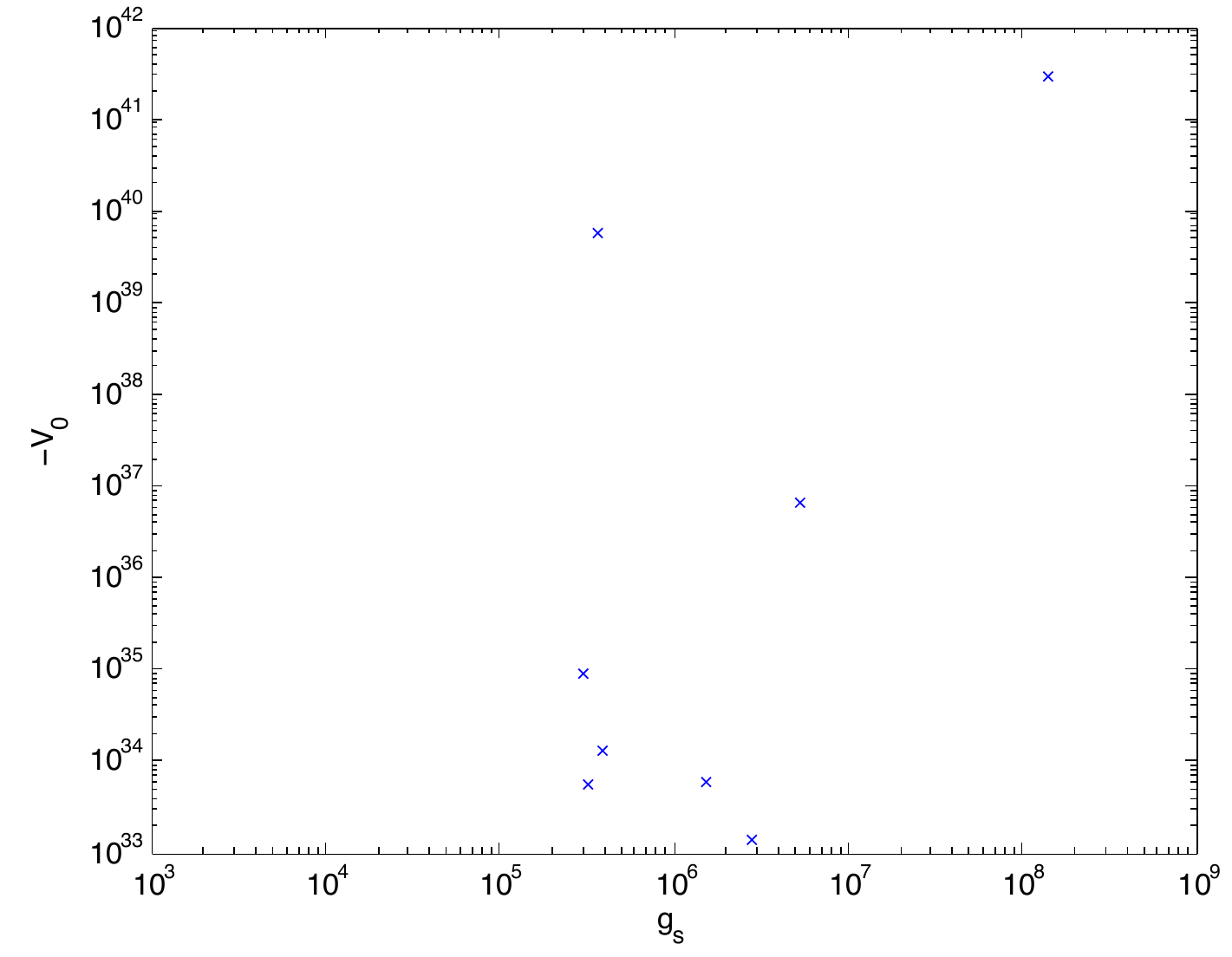} \includegraphics[width=7cm]{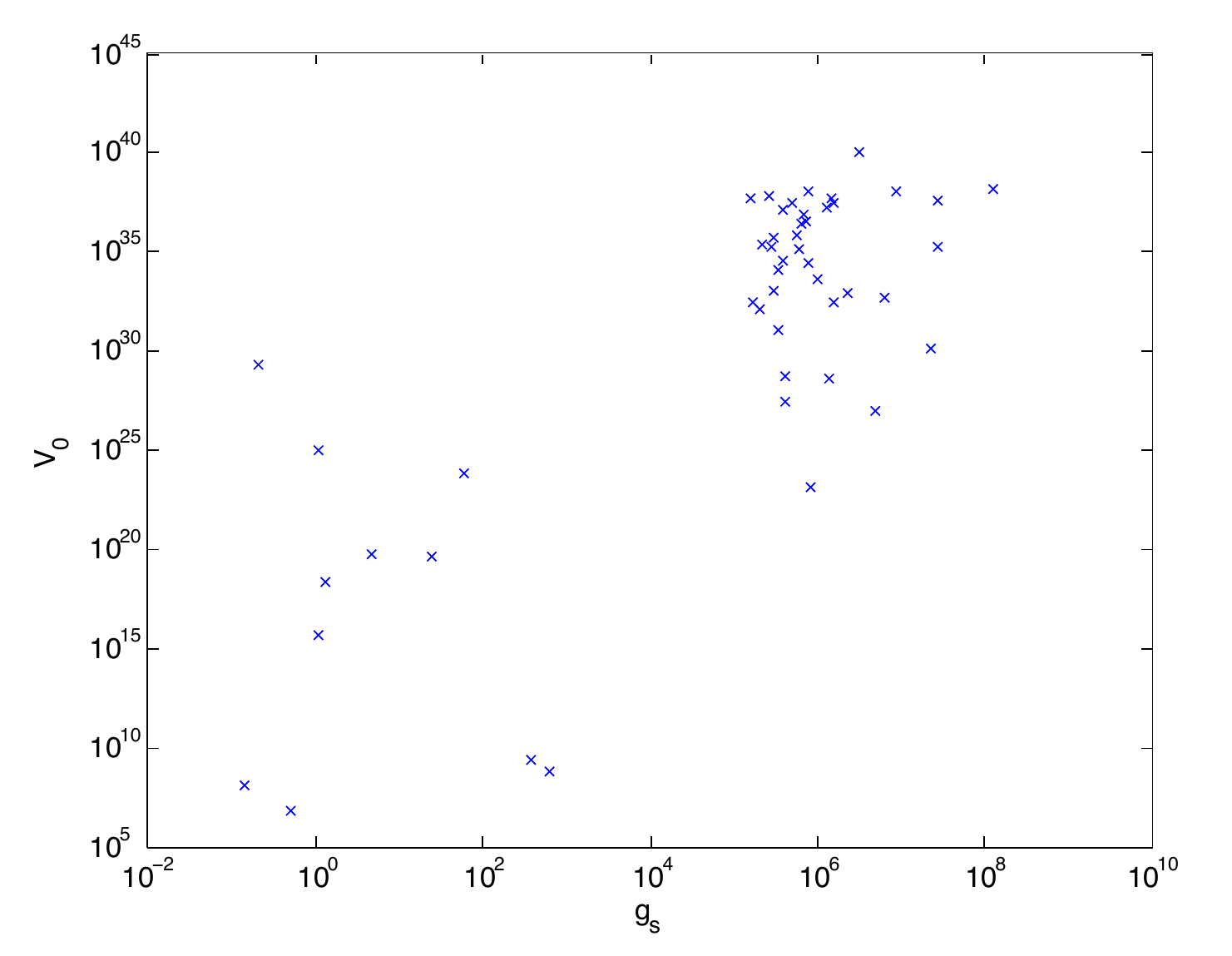}
\par\end{centering}
\caption{\label{land_tau} Landscape for non- SUSY vacua  with SUSY broken through the complex structure. (a) AdS, (b) dS.}
\end{figure}

\begin{figure}
\begin{centering}
\includegraphics[width=12cm]{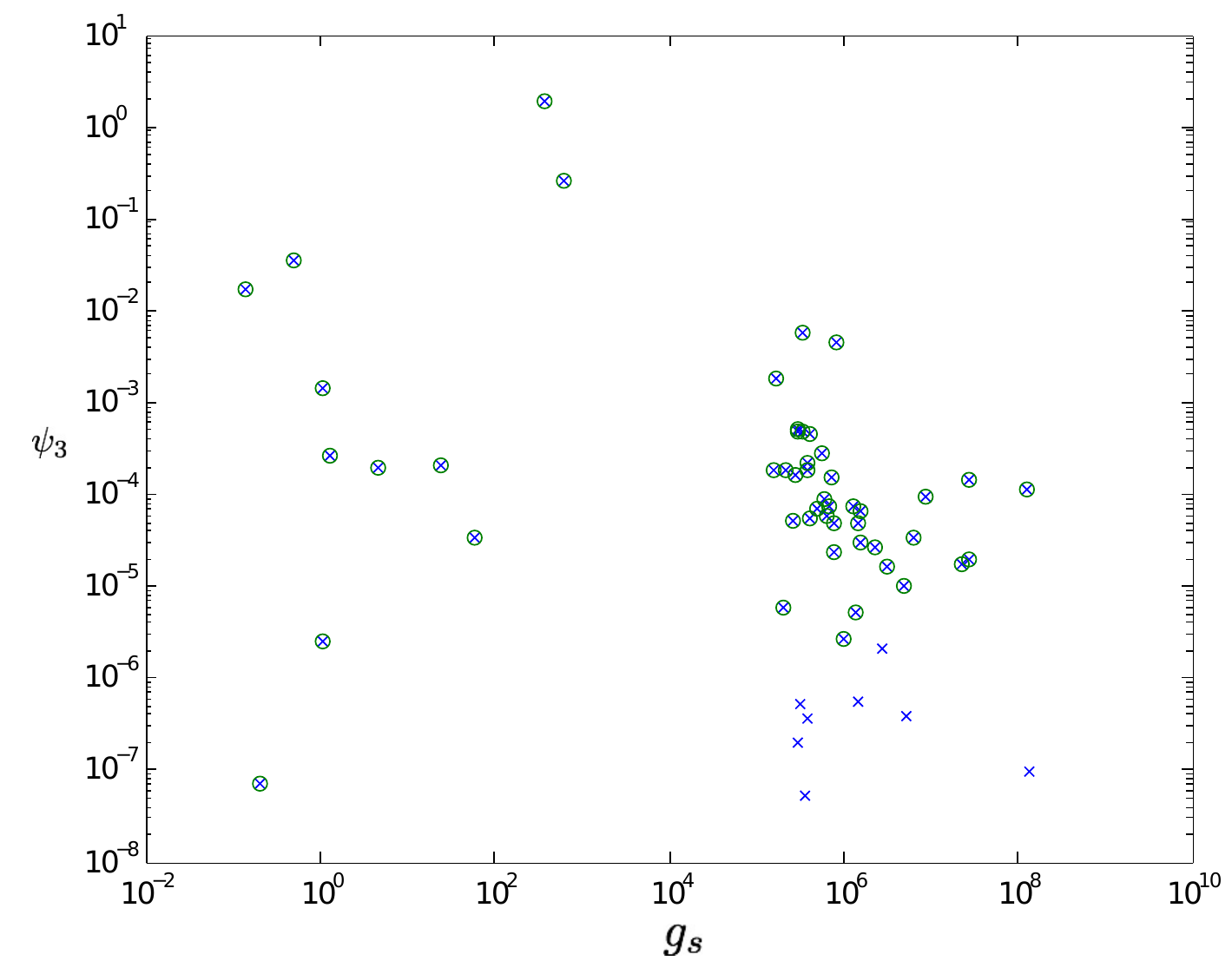}
\par\end{centering}
\caption{\label{coup_const_tau} Coupling constant versus volume of the internal space for all vacua. De Sitter vacua are marked as $\otimes$ while AdS vacua are marked as $\times$.}
\end{figure}

There are dS vacua solutions by breaking SUSY through the complex structure $\Phi_1$, as shown by the corresponding $\Lambda$-space distribution in Figure \ref{land_tau}. It can be seen that most of the vacua are dS and only few of them are AdS. In both cases the energy vacuum values are much bigger than expected and most important,  we observe that for all AdS vacua and almost all dS , the string coupling constant is very large, rendering our solutions unviable since they are far away from the perturbative regime where our approach is valid. Actually, there are only 3 cases for which $g_s$ is lesser than 1.
However,  the vacua distribution in terms of the internal volume (see Figure \ref{coup_const_tau}), lies in a region of small volume (small Kahler modulus) suggesting  that the supergravity approximation is not valid precisely for those cases with $g_s<1$. Therefore, for all found  vacua   in which SUSY is broken by the complex structure, there are not physical consistent conditions.\\

\subsection{Case 2: SUSY breakdown through all moduli}
Let us focus on  a particular subset of the moduli space by considering only the case in which supersymmetry is broken by all moduli in the absence of exotic orientifold planes, i.e., with $\tilde{N}=0$.  This forces us to consider only even fluxes. We shall consider the presence of exotic orientifolds at the end of the present section.\\

\begin{figure}
\begin{centering}
\includegraphics[width=7cm]{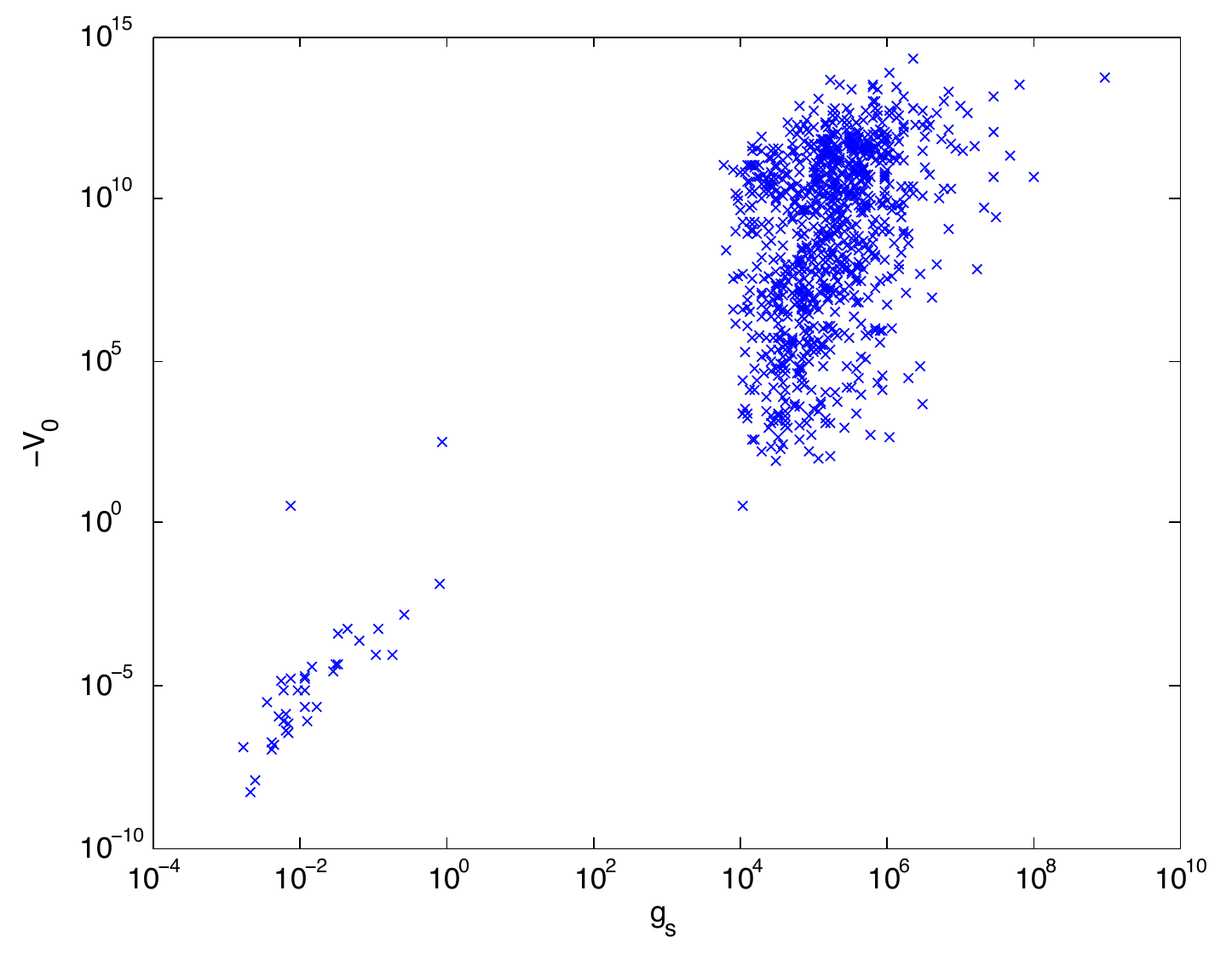} \includegraphics[width=7cm]{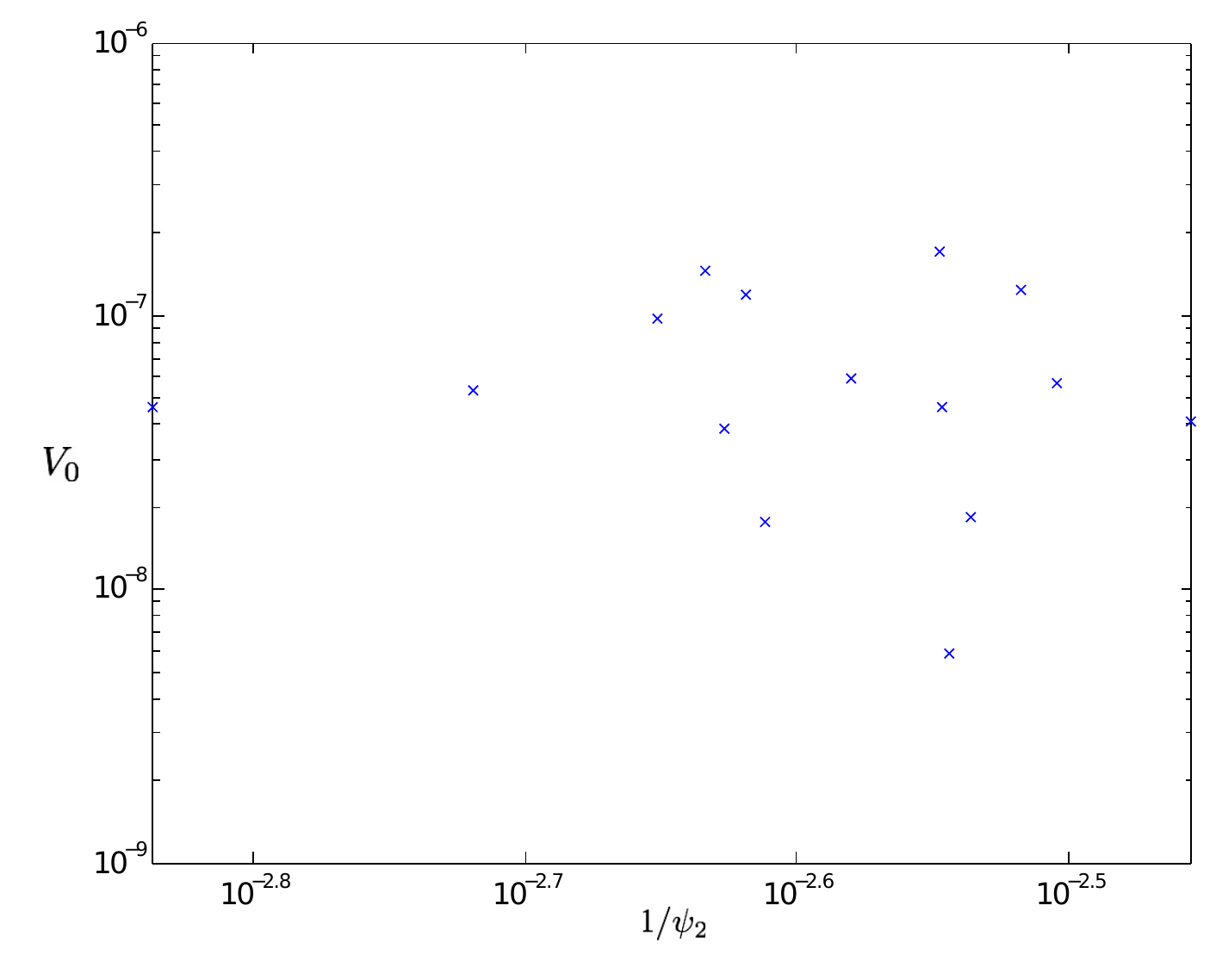}
\par\end{centering}
\caption{\label{landscape} Landscape for SUSY breaking through all the moduli: (a) AdS vacua, (b) dS vacua}
\end{figure}

After applying the genetic algorithm, we find two thousand vacua out of which only 15 have positive vacuum expectation values (vev's) and the rest are related to AdS vacua.
For negative energy values, the landscape is shown in Figure \ref{landscape}a. However we find that almost all physically attractive cases are unstable. This follows from the presence of  tachyonic directions in the eigenvalues of the mass matrix \cite{Abbott:1981ff, Gazeau:2006uy}
\begin{equation}
M^2_{ij}=D_i\partial_j\mathscr{V}+\frac{2}{3}V_0{\mathscr K}_{ij},
\label{massnd}
\end{equation}
where $D$ is a covariant-K\"ahler derivative and $V_0$ is the value of the scalar potential $\mathscr{V}$ at the minimum.  The number of stable AdS vacua is of order 10 with half of them presenting vev's of order $10^{-2}$. In table \ref{tab:fluxesstable} we show 5 representative cases.\\
\begin{center}
\begin{table}[h]
\caption{\label{tab:fluxesstable} Flux configuration  for stable AdS vacua.}
\centering
\begin{tabular}{c c c c c c c c}
\hline
\bf{Fluxes} & \bf{1} & \bf{2} & \bf{3} & \bf{4} & \bf{5} & \\
\hline

$a_{00}$	 				& 2		& 2		& 2  			& 2  			& 2 \\
$a_{01}$ 					& 16		& 16		& 18  		& 16 		& 20	\\
$a_{02}$ 					& 8		& 8		& 2  			& 14 		& 18	\\
$a_{03}$ 					& 4		& 34		& 20		  	& 24 		& 38\\
$a_{13}$					& 16		& 16		& 16 		& 16			& 16	\\
$a_{22}^1$ 				& 2		& 4  		& 2 	 		& 2 	 		& 4\\
$a_{23}$ 					& 16		& 32 	& 18 		& 16			& 40	\\
\hline
\end{tabular}
\end{table}
\end{center}
Concerning dS vacua, 
some of them correspond to solutions with a large string coupling constant $g_s$ where our approximations are not valid. The rest of them are related to large and small vev's. Focusing on those with small volumes, we report 10 different flux configurations leading to positive values of the corresponding classical scalar potential at its minimum, with small values of $g_s$ and small volumes\footnote{Notice that this does not mean that LARGE volume scenarios in presence of non-geometric fluxes are discarded.} (see  Figure \ref{landscape}b). Flux configurations for stable dS vacua with the above mentioned properties are listed in Table \ref{fig:dsfluxesstable}.\\
\begin{center}
\begin{table}[h]
\caption{\label{fig:dsfluxesstable} Flux configuration for  stable dS vacua.}
\centering
\begin{tabular}{c c c c c c c c c c c c c}
\hline
\bf{Fluxes} & \bf{1} & \bf{2} & \bf{3} & \bf{4} & \bf{5} & \bf{6} & \bf{7} & \bf{8} & \bf{9} & \bf{10} \\
\hline
$a_{00}$ 				& 2		& 2		& 2			& 2		& 2		& 2		& 2		& 2		& 2		& 2		 \\
$a_{01}$ 				& 20		& 16		& 18			& 18		& 18		& 20		& 20		& 20		& 18		& 16		\\
$a_{02}$ 				& 40		& 18		& 20			& 8		& 34		& 6		& 8		& 10		& 30		& 28		\\
$a_{03}$ 				& 38		& 16		& 36			& 20		& 8		& 8		& 2		& 36		& 40		& 2		\\
$a_{13}$				& 16		& 16		& 16			& 16		& 16		& 16		& 16		& 16		& 16		& 16		\\
$a_{22}^3$ 			& 2		& 2  		& 2  			& 2		& 2		& 2		& 2		& 2		& 2		& 2		\\
$a_{23}$ 				& 20		& 16 	& 18			& 18		& 18		& 20		& 20		& 20		& 18		& 16		\\
\hline
\end{tabular}
\end{table}
\end{center}
\begin{figure}
\begin{centering}
\includegraphics[width=12cm]{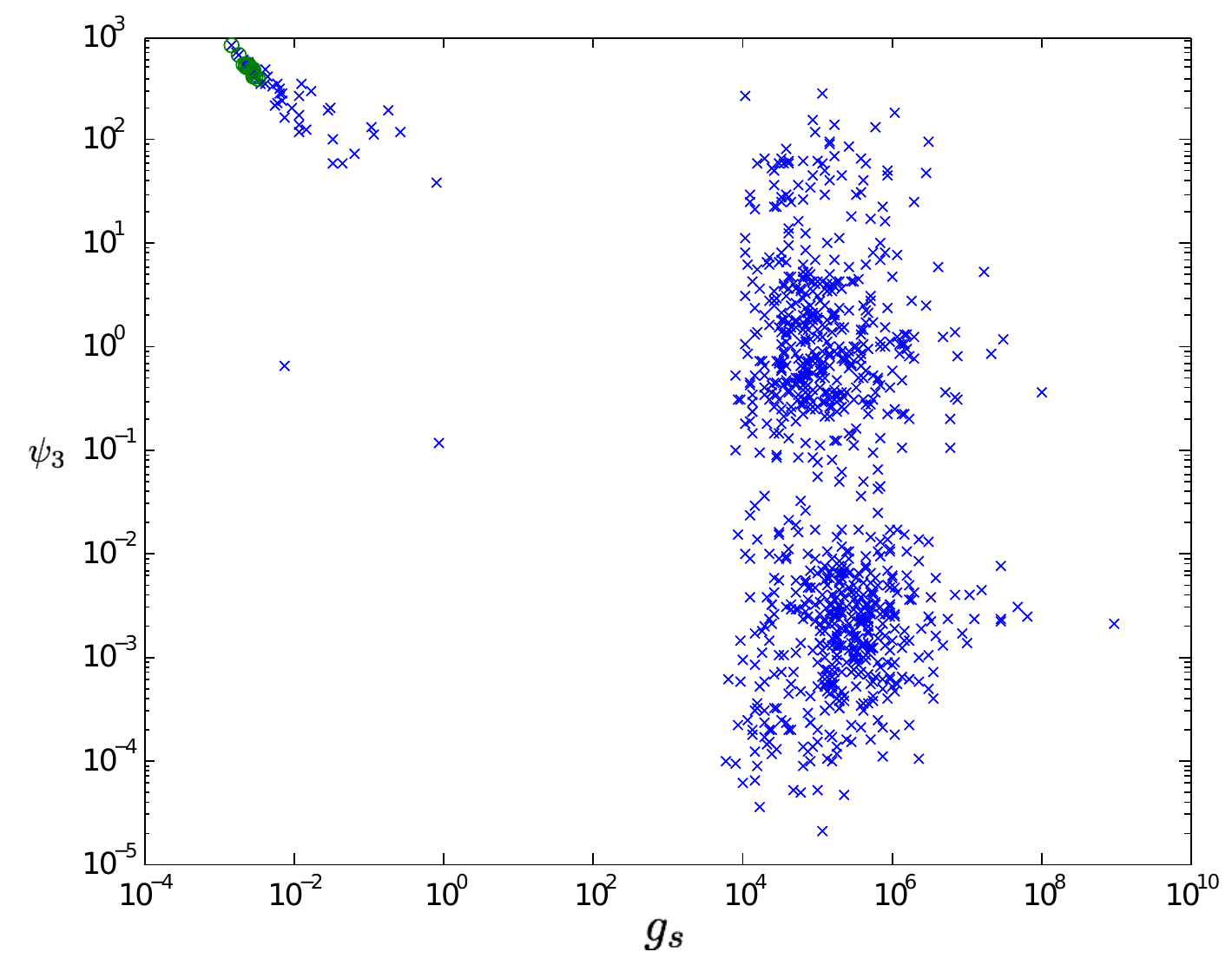}
\par\end{centering}
\caption{\label{coup_const} Coupling constant versus volume of the internal space.}
\end{figure}
From these calculations we observe that:
\begin{enumerate}
\item At least two non-geometric fluxes are required for the existence of stable (positive) vacua. We have not found a flux configuration with a single non-geometric flux leading to a stable $-$positive or negative valued$-$ vacua.
\item The order of stable dS and AdS vacua is the same ({\cal O}(1)).  As far as we know, this is the first case reported in which AdS is not favored against dS.\\
\end{enumerate}
Henceforth, within these 10 cases,  all moduli are stabilized
and their vacuum expectation values  are
shown in Table \ref{tab:mini}, together with the energy vacuum value which is of order $10^{-7}-10^{-8}$ in mass Planck units.
\begin{center}
\begin{table}[h]
\caption{\label{tab:mini} Expectation value of the moduli at the minimum.}
\centering
\begin{tabular}{c c c c c c c c}
\hline
\bf{Solution} & $\langle{\phi_1}\rangle$ & $\langle{\psi_1}\rangle$ & $\langle{\phi_2}\rangle$ & $\langle{\psi_2}\rangle$ & $\langle{\phi_3}\rangle$ & $\langle{\psi_3}\rangle$ & $V_{min}$\\
\hline 
\bf{1} 		& 0.01097& 0.0622 & 36.728 & 687.701 & 1.2790 & 854.393 & $1.7331{\times}10^{-8}$\\
\bf{2} 		& 0.0219 & 0.0661 & 50.4234 & 677.226 & 0.2078 & 861.222 & $1.7830{\times}10^{-8}$\\
\bf{3} 		& 0.0132 & 0.0530 & 64.6363 & 654.306 & 1.4210 & 856.003 & $1.2476{\times}10^{-8}$\\
\bf{4} 		& 0.0064 & 0.0702 & 23.0519 & 314.462 & 1.1425 & 416.45 & $1.6714{\times}10^{-7}$\\
\bf{5} 		& 0.0049 & 0.0735 & 30.5774 & 410.23 & 7.1069 & 512.971 & $7.0674{\times}10^{-8}$\\
\bf{6} 		& 0.0078 & 0.0561 & 89.8675 & 455.271 & 0.1171 & 555.596 & $7.1009{\times}10^{-8}$\\
\bf{7} 		& 0.0075 & 0.0639 & 16.9536 & 433.314 & 0.5245 & 501.969 & $9.8853{\times}10^{-8}$\\
\bf{8} 		& 0.0020 & 0.0700 & 15.2937 & 336.459 & 0.9326 & 432.311 & $1.7225{\times}10^{-7}$\\
\bf{9} 		& 0.0285 & 0.0658 & 65.7282 & 543.83 & 0.9524 & 673.968 & $8.0765{\times}10^{-8}$\\
\bf{10} 		& 0.0106 & 0.0691 & 11.5051 & 373.934 & 0.0187 & 499.693 & $6.4109{\times}10^{-8}$\\
\hline 
\end{tabular}
\end{table}
\end{center}
The moduli masses are obtained  by diagonalizing the mass matrix $M^2_{ij}$ from Eq.(\ref{massnd}) and their eigenvalues $\tilde{M}^2_{ij}$  are given presented in Table \ref{eigen},
where $\tilde{M}^2_{x}$ is the squared mass of the eigenvector $\tilde{X}$ constructed as a linear combination of the moduli fields $\phi_i$ and $\psi_i$ with $\tilde{X}=PX$ and $X^T=\{(\phi_i,\psi_i)\}^3_{i=1}=\{\chi_i\}_{i=1}^{6}$. The matrix $P$ which diagonalizes the  mass matrix $M^2$ by $\tilde{M}^2=P^{-1}M^2P$,  is given by
 \begin{equation}\label{eq:mateigen}
 P=
 \begin{bmatrix}
 -0.8854 & 0.4648 & O(10^{-7}) & O(10^{-5}) & O(10^{-6}) & O(10^{-5}) \\
 -0.4648 & 0.8854 & O(10^{-6}) & O(10^{-5}) & O(10^{-5}) & O(10^{-6}) \\
 O(10^{-6}) & O(10^{-5}) & -0.1439 & 0.0673 & 0.9855 & 0.0577 \\
 O(10^{-5}) & O(10^{-6}) & 0.0094 & 0.2361 & -0.0715 & 0.9690 \\
 O(10^{-5}) & O(10^{-5}) & 0.4191 & 0.8813 & 0.0137 & -0.2177 \\
 O(10^{-5}) & O(10^{-6}) & 0.8964 & -0.4036 & 0.1526 & 0.1009
 \end{bmatrix},
 \end{equation}
\begin{center}
\begin{table}[h]
\caption{\label{eigen} Mass Eigenvalues.}
\centering
\begin{tabular}{c c c c c c c}
\hline
\bf{Solution} & $\tilde{M}_{\phi_1}^2$ & $\tilde{M}_{\psi_1}^2$ & $\tilde{M}_{\phi_2}^2$ & $\tilde{M}_{\psi_2}^2$ & $\tilde{M}_{\phi_3}^2$ & $\tilde{M}_{\psi_3}^2$ \\
 \hline
\bf{1} & $6.354{\times}10^{-5}$ &  $1.251{\times}10^{-6}$ &  $1.153{\times}10^{-8}$ & $1.165{\times}10^{-8}$ & $1.153{\times}10^{-8}$ & $1.154{\times}10^{-8}$ \\
\bf{2} & $1.110{\times}10^{-4}$ &  $1.908{\times}10^{-6}$ & 	$1.179{\times}10^{-8}$ & $1.191{\times}10^{-8}$ & $1.179{\times}10^{-8}$ & 	$1.180{\times}10^{-8}$ \\
\bf{3} & $8.340{\times}10^{-5}$ & 	$1.235{\times}10^{-6}$ & 	$8.340{\times}10^{-9}$ &	 $8.459{\times}10^{-9}$ & $8.339{\times}10^{-9}$ & 	$8.341{\times}10^{-9}$ \\
\bf{4} & $2.268{\times}10^{-4}$ & 	$5.765{\times}10^{-6}$ & 	$1.112{\times}10^{-7}$ &	 $1.130{\times}10^{-7}$ & $1.112{\times}10^{-7}$ & 	$1.112{\times}10^{-7}$ \\
\bf{5} & $1.147{\times}10^{-4}$ & 	$3.764{\times}10^{-6}$ & 	$4.695{\times}10^{-8}$ &	 $4.798{\times}10^{-8}$ & $4.695{\times}10^{-8}$ & 	$4.697{\times}10^{-8}$ \\
\bf{6} & $1.505{\times}10^{-4}$ & 	$2.359{\times}10^{-6}$ & 	$4.734{\times}10^{-8}$ &	 $4.786{\times}10^{-8}$ & $4.734{\times}10^{-8}$ & 	$4.735{\times}10^{-8}$ \\
\bf{7} & $1.663{\times}10^{-4}$ & 	$3.266{\times}10^{-6}$ & 	$6.572{\times}10^{-8}$ &	 $6.662{\times}10^{-8}$ & $6.579{\times}10^{-8}$ & 	$6.573{\times}10^{-8}$ \\
\bf{8} & $2.175{\times}10^{-4}$ & 	$4.242{\times}10^{-6}$ & 	$1.147{\times}10^{-7}$ &	 $1.161{\times}10^{-7}$ & $1.478{\times}10^{-8}$ & 	$1.1478{\times}10^{-8}$ \\
\bf{9} & $2.533{\times}10^{-4}$ & 	$3.758{\times}10^{-6}$ & 	$5.377{\times}10^{-8}$ &	 $5.411{\times}10^{-8}$ & $5.377{\times}10^{-8}$ & 	$3.778{\times}10^{-8}$  \\
\bf{10} & $2.044{\times}10^{-4}$ & 	$4.579{\times}10^{-6}$ & 	$4.270{\times}10^{-8}$ &	 $4.389{\times}10^{-8}$ & $4.269{\times}10^{-8}$ & 	$4.271{\times}10^{-8}$, \\
\hline
\end{tabular}
\end{table}
\end{center}
with a real redefined moduli field $\tilde\chi_i $ given by
\begin{equation}
\tilde\chi_i=P_{ij}\chi_i.
\end{equation}
The corresponding fields  are approximately separated into two sets $(\tilde\phi_1, \tilde\psi_1)$ and $(\tilde\phi_i,\tilde\psi_i)$ with $i=2,3$. For $i=1,2$ we observe that  $j=1,2$, while for $i=3,\dots ,6$, the index $j$ run  over $\{3,\dots 6\}$. If we consider  the fact that the masses for the  complex structure moduli are much larger then clearly we have a hierarchy on masses. It seems that once we find stable dS vacua, all of them present such hierarchy. We shall extensively use this feature in the next section  to find suitable conditions for inflation.\\

\subsection{S-goldstino direction}

We are now ready to compute which modulus\footnote{From now on, we shall refer to the redefined moduli fields $\widetilde\Phi$  as $\Phi$.} actually breaks SUSY: consider that for a particular set of  vev's for all the moduli, we have a SUSY configuration, implying that all K\"ahler derivatives vanish for all the moduli. By departing from such values, SUSY is broken,  rendering some of the F-terms to be different from zero. Since the F-terms are given by
\begin{equation}
<F^i>=<e^{\mathscr{K}/2}{\cal D}_{\bar{i}} \mathscr{W}\mathscr{K}^{i\bar{i}}>,
\end{equation}
with the K\"ahler potential $\mathscr{K}$ being diagonal, this only happens if $<F^1>$ becomes non-zero while the other are kept null, otherwise, as we have seen, Tadpole constraints and Bianchi identities are not fulfilled. Therefore, we conclude that the complex structure breaks SUSY. This is confirmed by computing the F-terms which for the complex field $\Phi_{2,3}$ are of order $10^{-5}$ while $F^1$ is of order $10^3$. \\

In the same line of reasoning, the gravitino mass
\begin{equation}
m_{3/2}=e^{\mathscr{K}/2}W,
\end{equation}
can be evaluated at the minimum of the scalar potential.  Table \ref{tab:massgravitino} shows the values for the gravitino mass and the $F^1$ term vev.
\begin{center}
\begin{table}[h]
\caption{\label{tab:massgravitino} Gravitino mass and SUSY breaking scale.}
\centering
\begin{tabular}{p{2cm} p{0.65cm} p{0.65cm} p{0.65cm} p{0.65cm} p{0.65cm} p{0.65cm} p{0.65cm} p{0.65cm} p{0.65cm} p{0.65cm}}
\hline
 & \bf{1} & \bf{2} & \bf{3} & \bf{4} & \bf{5} & \bf{6} & \bf{7} & \bf{8} & \bf{9} & \bf{10} \\
\hline
\small{\bf{$m_{3/2} ({\times}10^{-4}$)}}& \small{$1.796$}	& \small{$2.143$}	& \small{$1.701$}	&  \small{$3.217$} & \small{$2.282$} & \small{$2.049$} & \small{$2.449$} & \small{$2.293$} & \small{$3.017$} & \small{$3.028$}  \\
\small{\bf{$F^1$-term}} & \small{$577.29$}	& \small{$678.42$}	& \small{$490.70$}   	& \small{$308.75$} & \small{$376.77$} & \small{$333.16$} & \small{$349.25$} & \small{$321.70$} & \small{$664.50$} & \small{$352.33$}  \\
\hline
\end{tabular}
\end{table}
\end{center}
We see that:  
\begin{enumerate}
\item The s-goldstino direction given by the F-terms, is approximately given by $F^1$ indicating that it points towards the K\"ahler derivative with respect to $\Phi_1$ ( i.e., the s-goldstino direction  is identified with the complex structure direction). 
\item The gravitino mass and the scale at which SUSY is broken present very high values, around $10^{14}$ GeV and $10^{21}$ GeV, respectively. They are still far from being realistic. We  briefly comment on this in section  5.
\end{enumerate}
Our next goal is to study if there are conditions for slow-roll inflation for small and/or large fields and determine if there is a single-field or a multi-field scenario for inflation. We also want   to see if there is a preferential direction  for inflation with respect the s-goldstino.

\section{Non-geometric slow-roll Inflation}
The presence of a hierarchy on the masses suggests a natural candidate for the inflaton,  either composed by the complex structure $\Phi_1$ or by a combination of the dilaton $\Phi_2$ and the K\"ahler moduli $\Phi_3$. In order to establish  which case is supported by our calculations we must consider small or large field scenarios. Since the last yields to a proper construction of a stable inflationary trajectory, which is quite complicated to find, we shall focus our analysis to small fields and leave some qualitative comments on the large-field regime after numerically computing the slow-roll parameters.\\

In the small-field scenario, the complex structure $\Phi_1$ cannot be identified with the inflaton since its mass at the minima is much larger than the Hubble constant $V_0$. This leads  to a possible multi-field inflation composed by $\Phi_2$ and $\Phi_3$.  In such case, the complex structure would decay\footnote{Since we are not considering the presence of D-branes, the decay of the moduli fields via the interaction with other forces is absent in our model. The reader must understand this point as the possibility to decay if coupling terms in an effective action would be present.} before the rest of moduli by which the goldstino would play the role of a Polonyi field\footnote{We do not have so far a solution for the cosmological moduli problem \cite{Coughlan:1983ci, Banks:1995dt}.}. Notice as well that
at this point there is still the possibility that inflation is driven by a single real field. We shall shortly see that this is not the case.\\

\subsection{Slow-roll inflation}
Despite the generalization to multi-fields it is still possible to have slow-roll inflation by defining an appropriate parameter that must be smaller than one. These are given by:
\begin{eqnarray}
\epsilon_T&=&\sum_i\epsilon_i=\sum_i\left(\frac{\partial_i V}{V}\right)^2<<1,\\
\eta_{ij}&=&\frac{D_i\partial_j V}{V}<<1.
\end{eqnarray}
More over, there must be conditions on which inflation ends,  reflected as conditions upon which the slow-roll conditions are violated. \\

\begin{figure}
\begin{centering}
a) \includegraphics[height=5.cm]{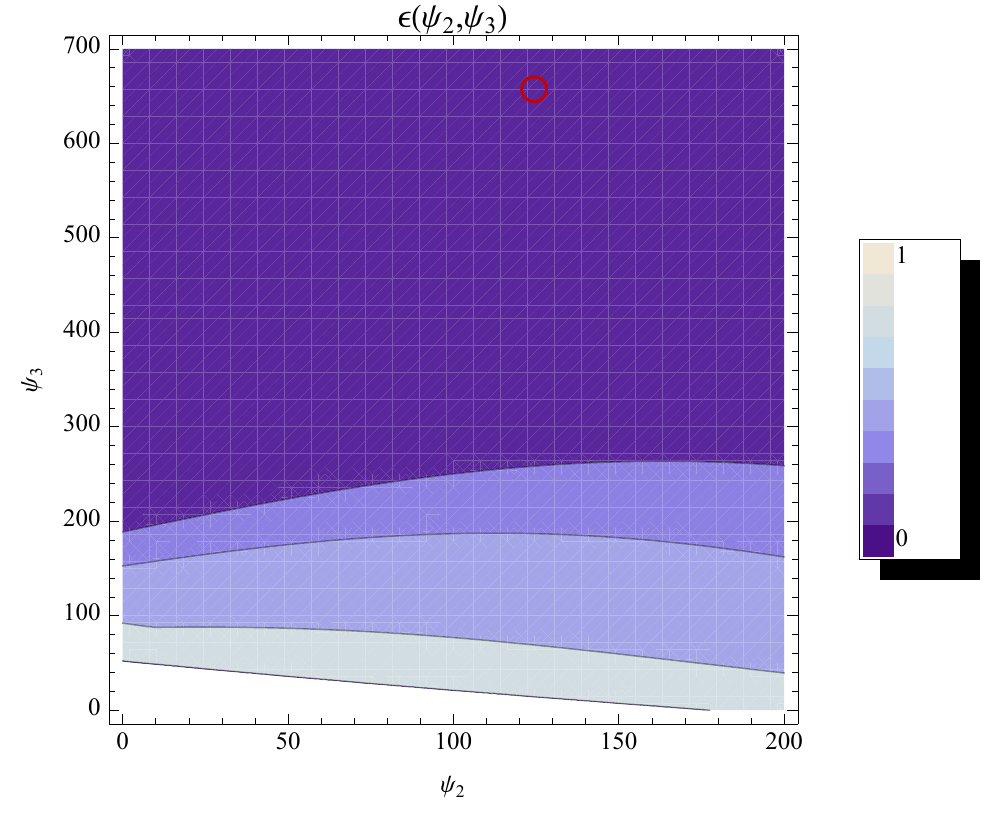} b) \includegraphics[height=5.cm]{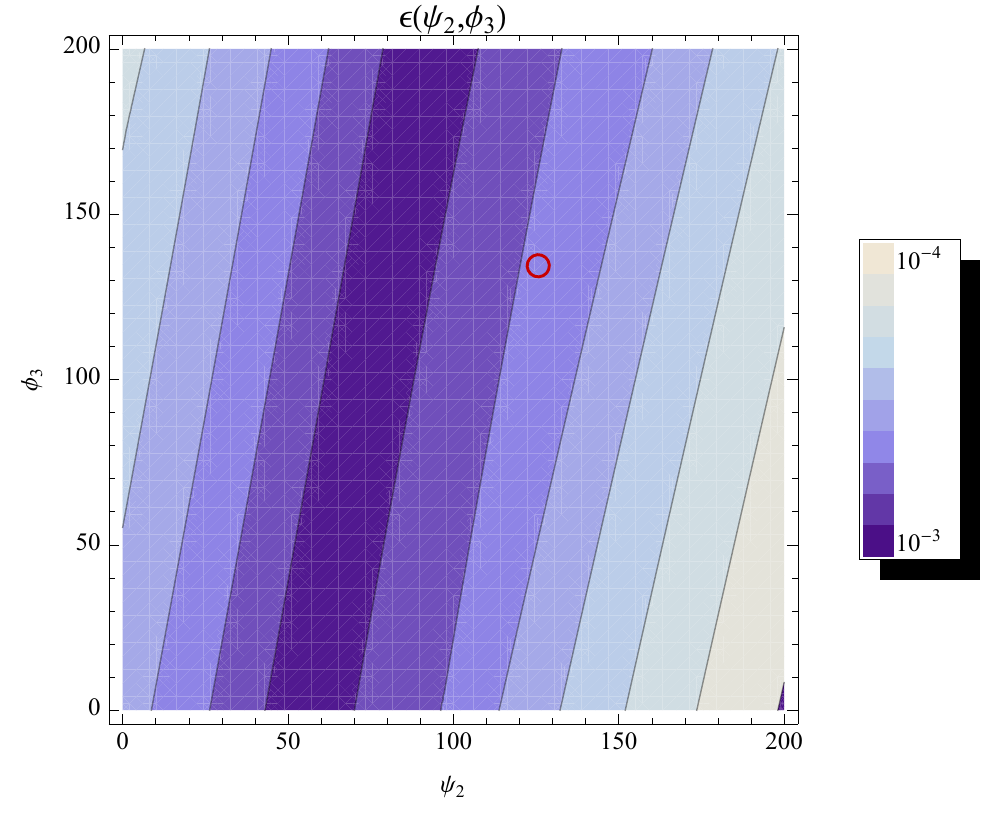}\\
c) \includegraphics[height=5.cm]{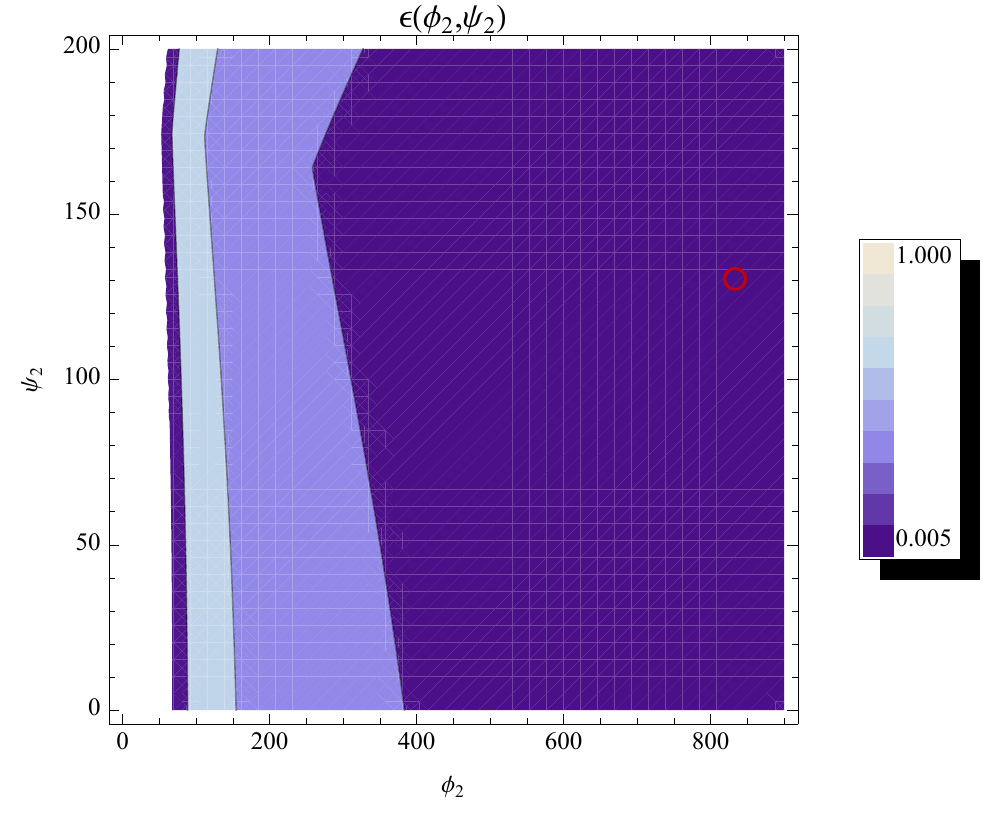} d) \includegraphics[height=5.cm]{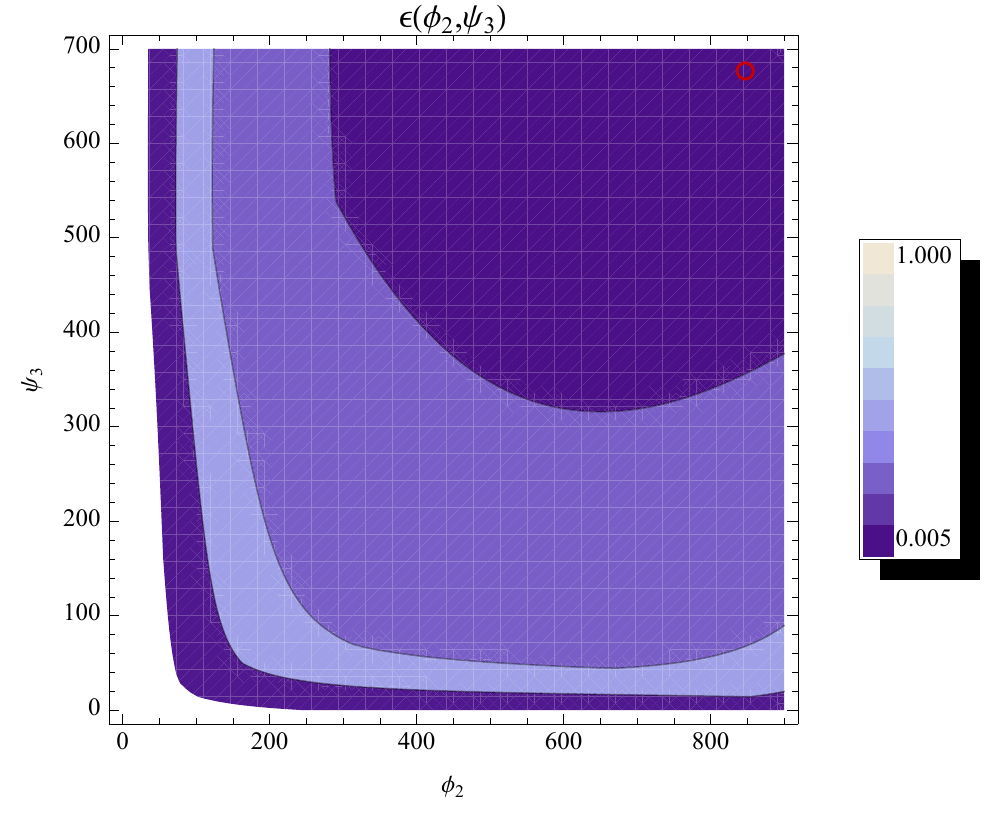}\\
e) \includegraphics[height=5.cm]{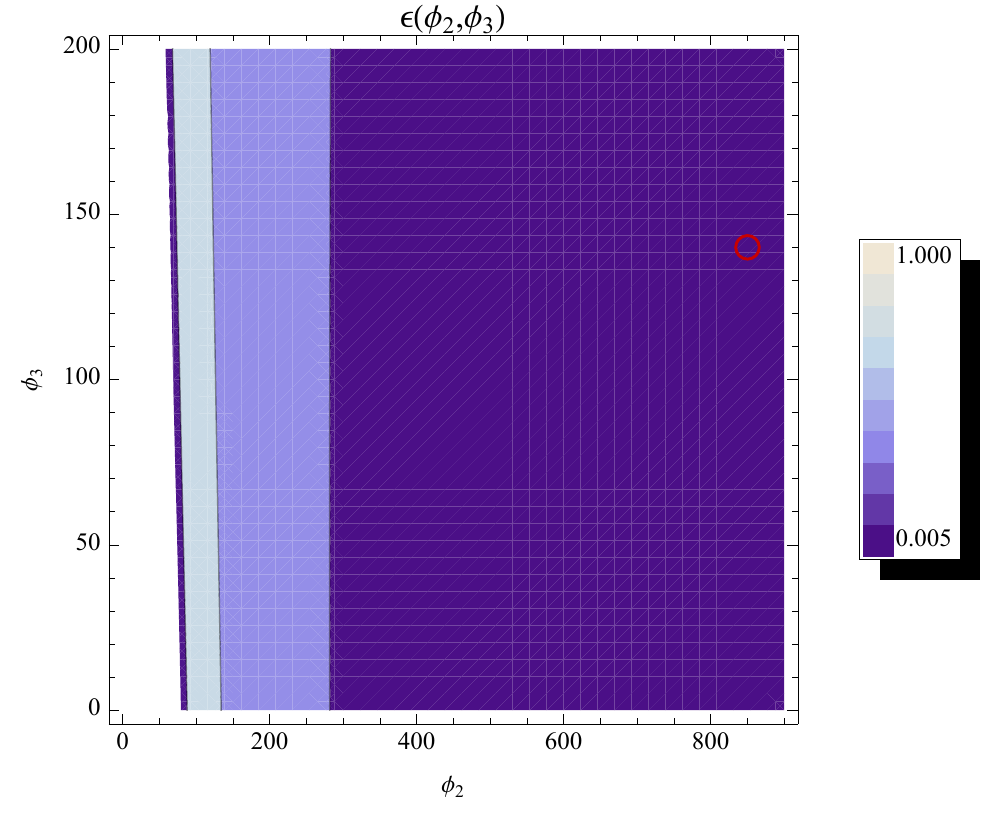} f) \includegraphics[height=5.cm]{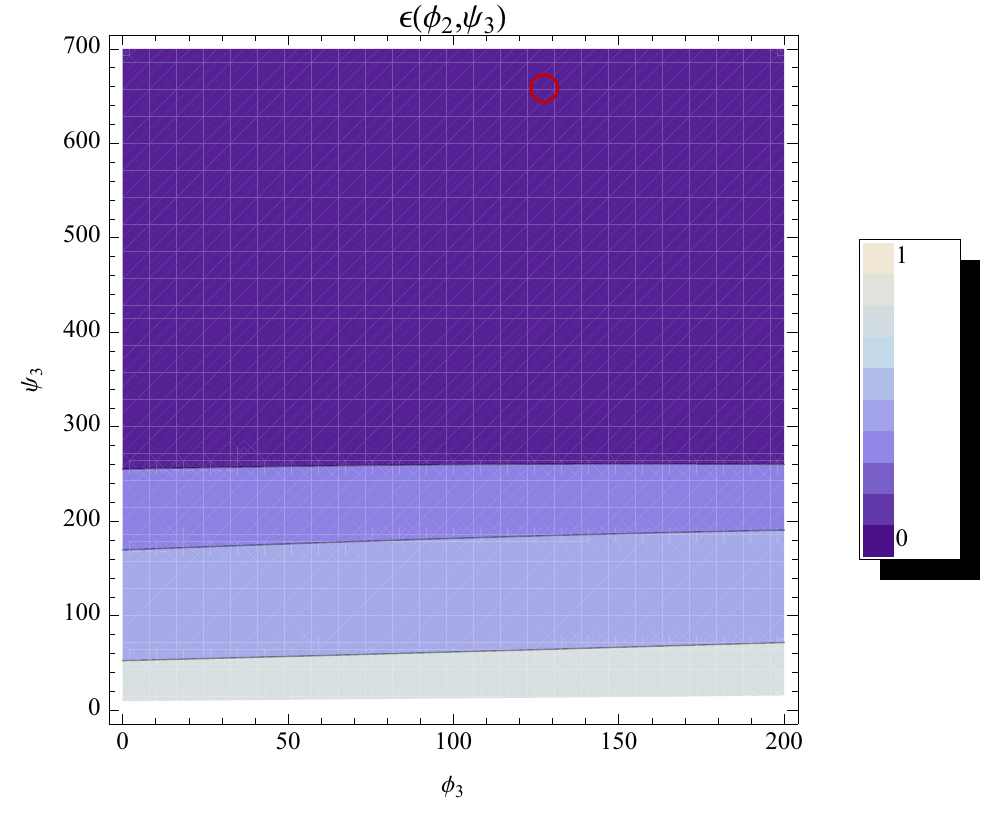}
\par\end{centering}
\caption{\label{fig:epsilon}  Projections of parameter $\epsilon$ in the plane a) $\psi_2$-$\psi_3$ b)  $\psi_2$-$\phi_3$ c)  $\phi_2$-$\psi_2$ d)  $\phi_2$-$\psi_3$ e)  $\phi_2$-$\phi_3$ f)  $\phi_3$-$\psi_3$. The rest two real moduli are kept constant at their vev. The red circle denotes the position of the vev's for the two real fields.}
\end{figure}

In this case, the slow-roll parameters depend on 4 real fields and finding explicit and stable inflationary directions seems to be a difficult task.  Instead we look for simple conditions for inflation to be present by projecting the slow-roll parameters in 2-dimensional planes where 2 real moduli are kept constant. If inflationary directions do not appear in such projections we could not conclude that multi-field inflation is absent. But if we are able to find some inflationary directions in these two-dimensional projections we can assure that at least we have a subset of all possible inflationary trajectories. Projections for  the $\epsilon$ parameter are shown in Figure \ref{fig:epsilon},   where the two of the moduli are taken constant at their stabilized values. Similar plots are obtained by keeping two of the moduli at constant values away from the minimum of the potential.\\

From the above figures we conclude the following:
\begin{enumerate}
\item The scalar potential $\mathscr{V}$ is quite flat with respect to $\phi_3$ and $\psi_2$ while the others are kept constant.
\item There are important variations on the $\eta$ parameter as $\phi_2$ and $\psi_3$ moves from the minima. 
\item It is possible to qualitatively notice the existence of inflationary trajectories on which inflation ends by facing zones where $\epsilon$ reaches values of order 1.
\item It seems that a combinations of $\phi_2$ and $\psi_3$ would be a representative $\epsilon$-inflationary trajectory.
\item Since $\psi_3$ parametrizes the internal volume, we see that for an inflationary trajectory driven by variations of $\psi_3$,  far from the minima (still in the small-field scenario), the internal volume should be very small and it increases its size as it runs towards the minima of the potential. Therefore, inflation  would be driven as the internal space increases its size. 
\end{enumerate}

\begin{figure}
\begin{centering}
a) \includegraphics[height=5.cm]{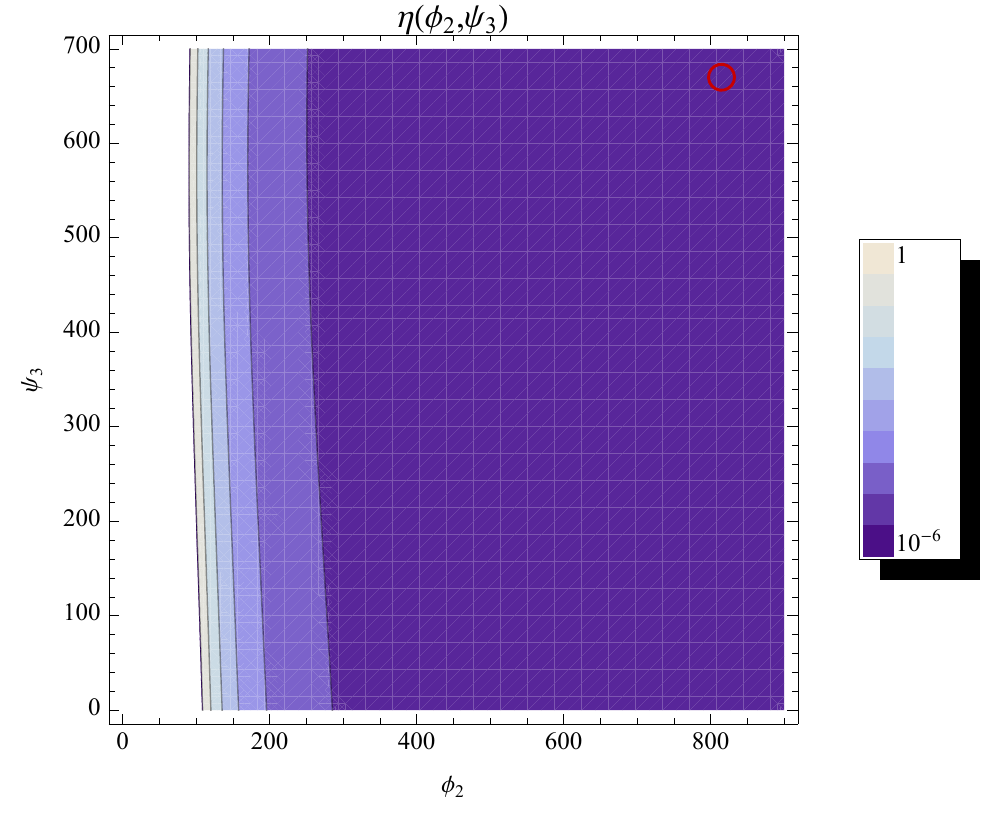} b) \includegraphics[height=5.cm]{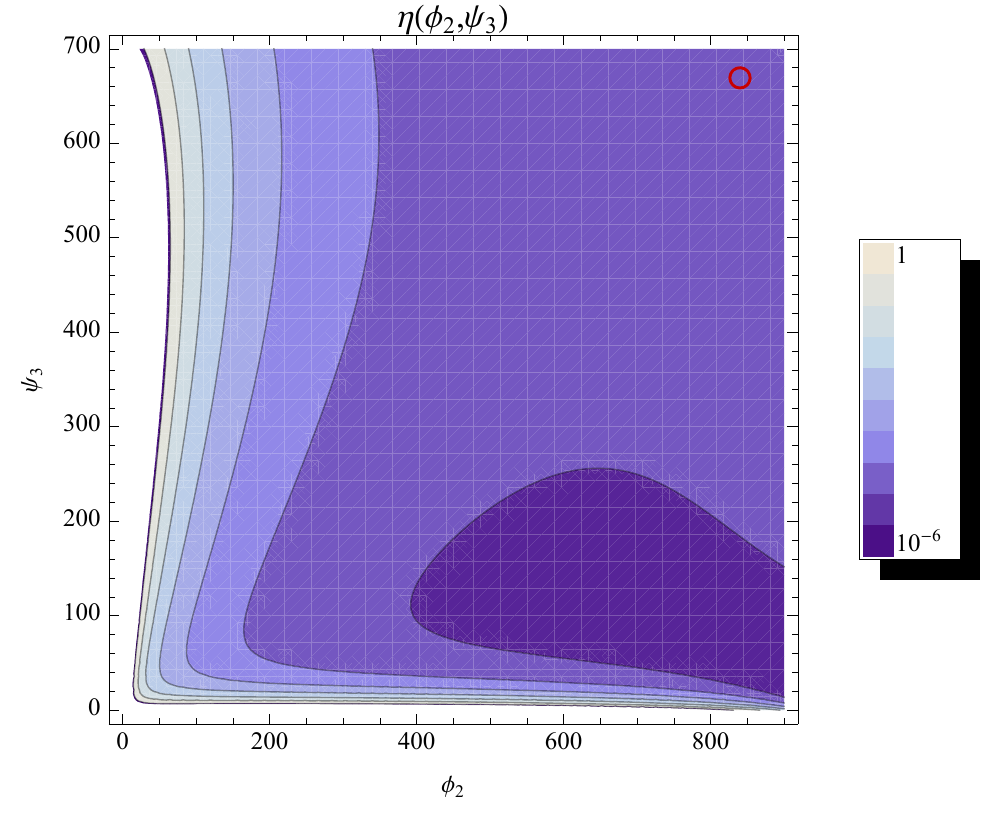}\\
c) \includegraphics[height=5.cm]{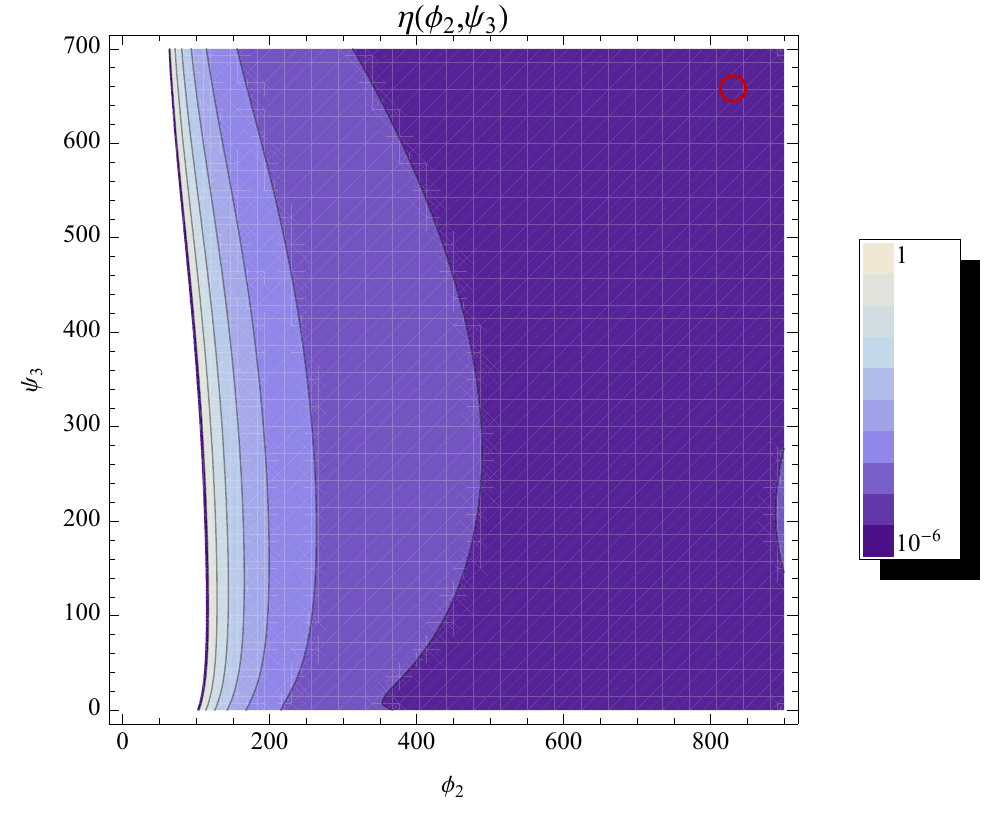} d) \includegraphics[height=5.cm]{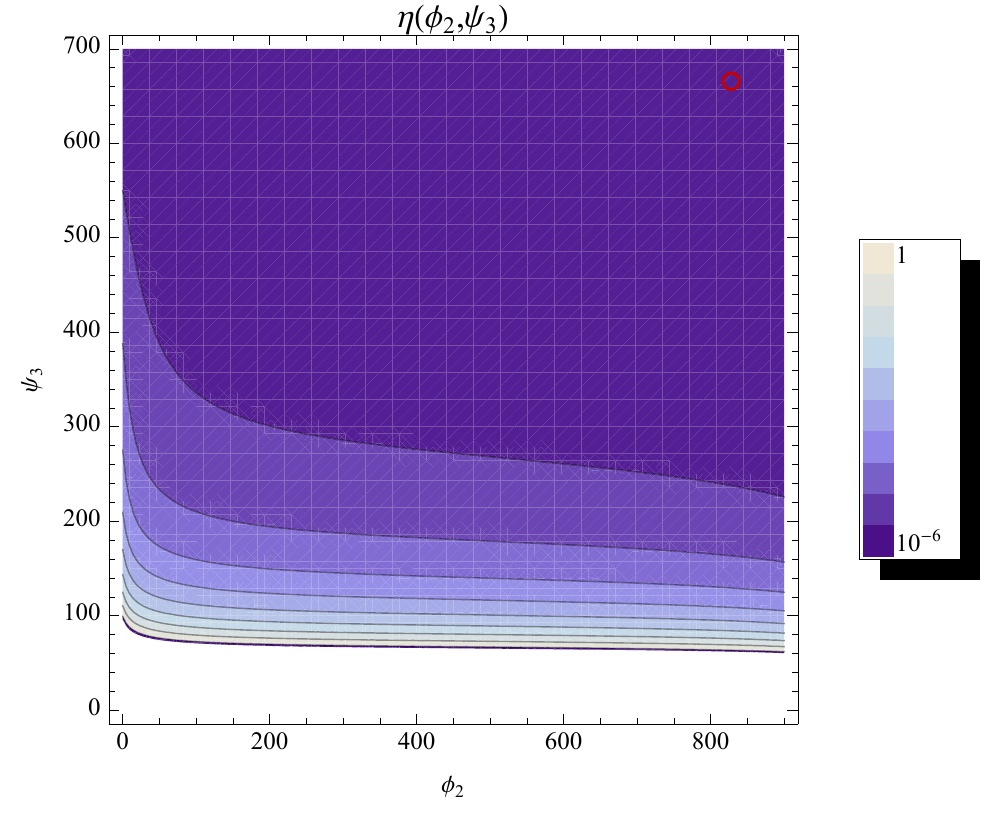}
\par\end{centering}
\caption{\label{fig:eta}  Projections of the matrix elements $\eta_{ij}$ on the space $\phi_2$-$\psi_3$. a) $\eta_{33}$, b)$\eta_{44}$, c)$\eta_{55}$, d)$\eta_{66}$.}
\end{figure}

Let us now project the matrix terms $\eta_{ij}$ in  two-dimensional planes and see if there are still inflationary directions in the field space.  We select the matrix elements concerning only the fields $\phi_2$ and $\psi_3$ since we have already see that $\epsilon$-inflationary trajectories exist. The projections are shown in Figure \ref{fig:eta}, from which we observe that there are conditions for inflation to exist and qualitatively trajectories for inflation to end.\\

\begin{figure}
\begin{centering}
a) \includegraphics[height=5.cm]{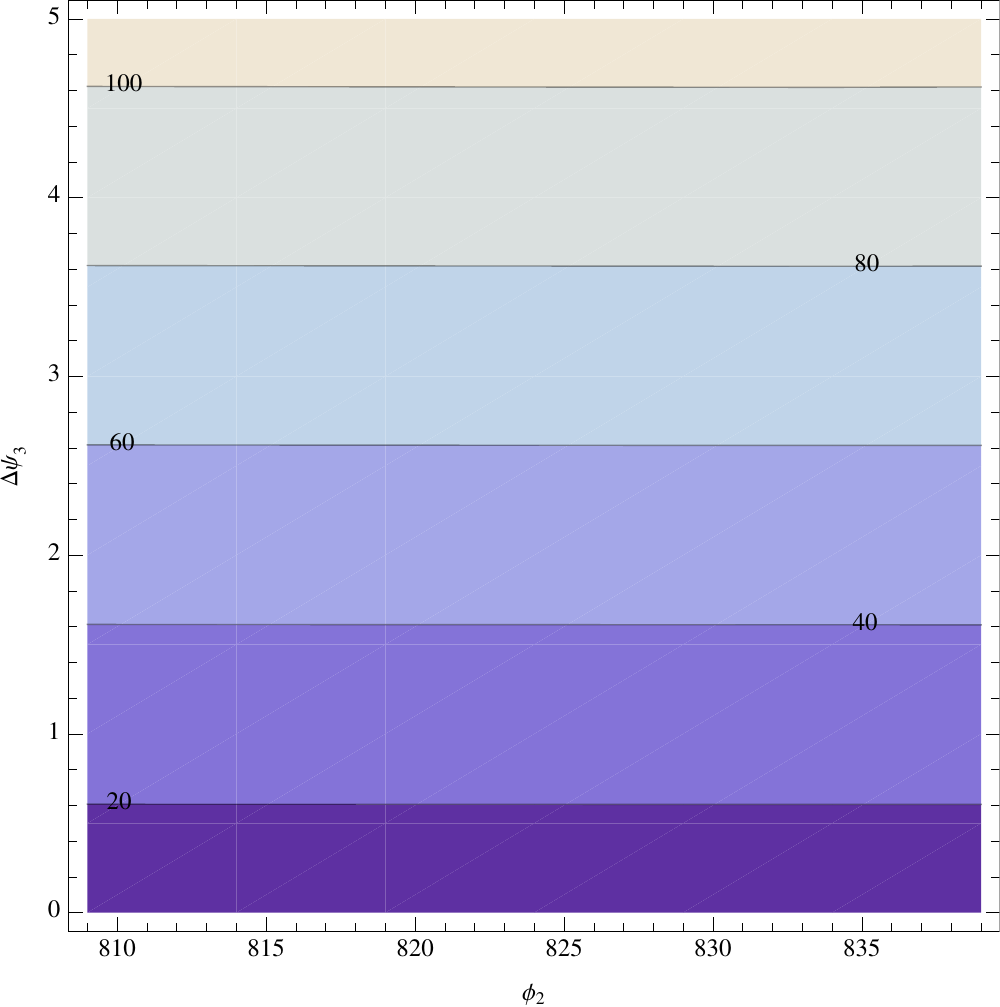} b) \includegraphics[height=5.cm]{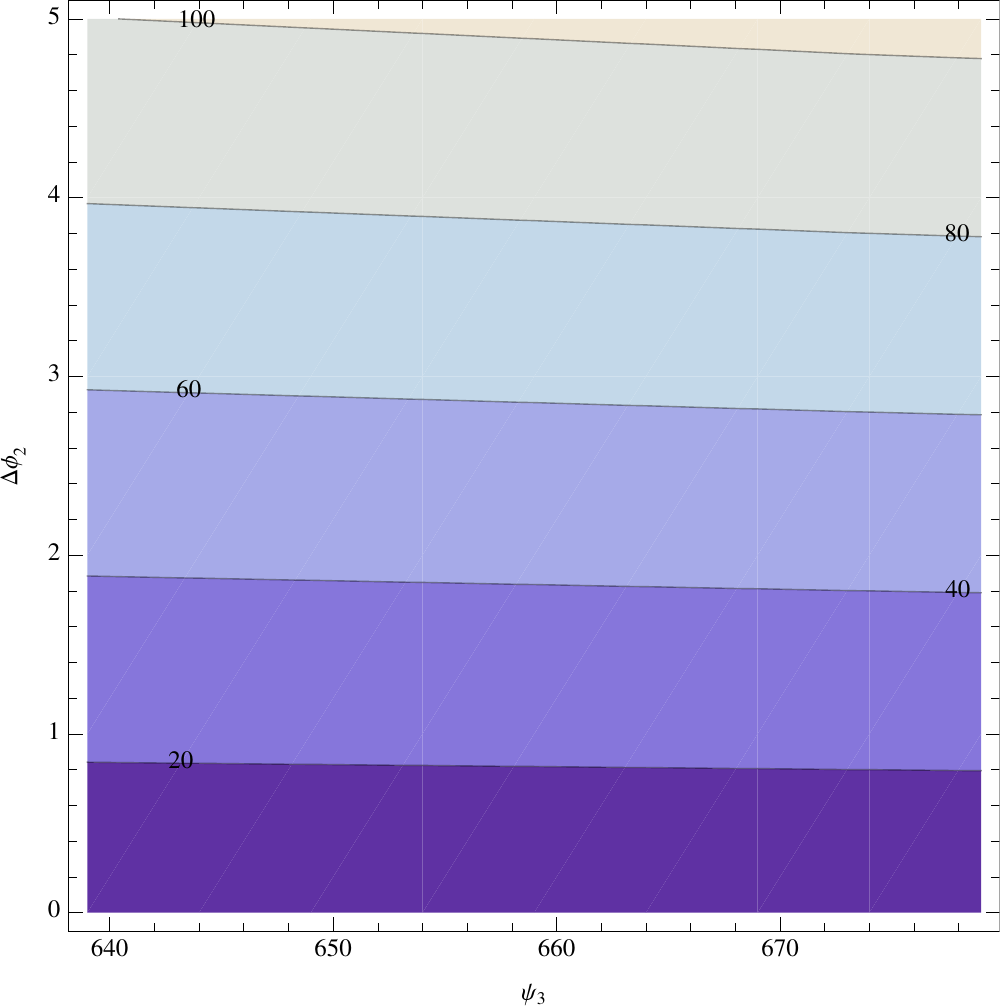}\\
\par\end{centering}
\caption{\label{fig:e-folds}  a) $\Delta{\psi_3}$ {\it vs} $\langle{\phi_2}\rangle$ b) $\Delta{\psi_2}$ {\it vs} $\langle{\psi_3}\rangle$.}
\end{figure}

In summary, we have found that there are suitable stable inflationary directions. Particularly we have found that in the  $\phi_2$-$\psi_3$ plane such directions exist. This leads us to the conclusion that at least for this case, the inflaton direction is (approximately) orthogonal to the s-goldstino. However, we should still show that inflation is sufficiently long to produce 60 e-folds. At this point this can be achieved either by a two-field inflation or by a single-field inflation, since we observe from Figure \ref{fig:epsilon} that keeping constant one of the two real moduli $\phi_2$ and $\psi_3$ we still have inflationary trajectories. The number of e-folds required from phenomenological considerations to end inflation is given by:
\begin{equation}
 N(\tilde\phi)=\int_{{\tilde\phi_{end}}}^{\tilde\phi} \frac{d\tilde\phi}{\sqrt{2 \epsilon_V(\tilde\phi)}}\sim 60,
\end{equation}
where $\tilde{\phi}$ is a linear combination of the moduli fields $\phi_2$ and $\psi_3$.  In Figure \ref{fig:e-folds} we show the plot of the e-folding values by keeping constant one of the two inflationary directions.
As it is shown,  it is impossible to cover the 60 e-foldings by small perturbations of the field, either by keeping constant $\phi_2$ or $\psi_3$. This is reflected from the fact that single-field trajectories joining the minimum with the region in which $\eta\sim 1$, are sufficiently short to produce the required number of e-foldings. In other words, we find that single field inflation in the small-field scenario is discarded. This result is in agreement with the bounds computed in \cite{Borghese:2012yu}.\\

On the other hand, 60 e-folds can be achieved by many possible trajectories in the two-field inflation scenario, as shown in Figure \ref{fig:epsilon}, where a trajectory joining the minimum of the potential to a region in which $\epsilon \sim 1$ can be constructed by varying the two fields. It is still necessary to have a more detailed description about the explicit construction of such trajectories as well as a dynamical analysis about their stability.

\subsection{Stable vacua with exotic orientifolds}
\begin{figure}
\begin{centering}
a) \includegraphics[height=5.cm]{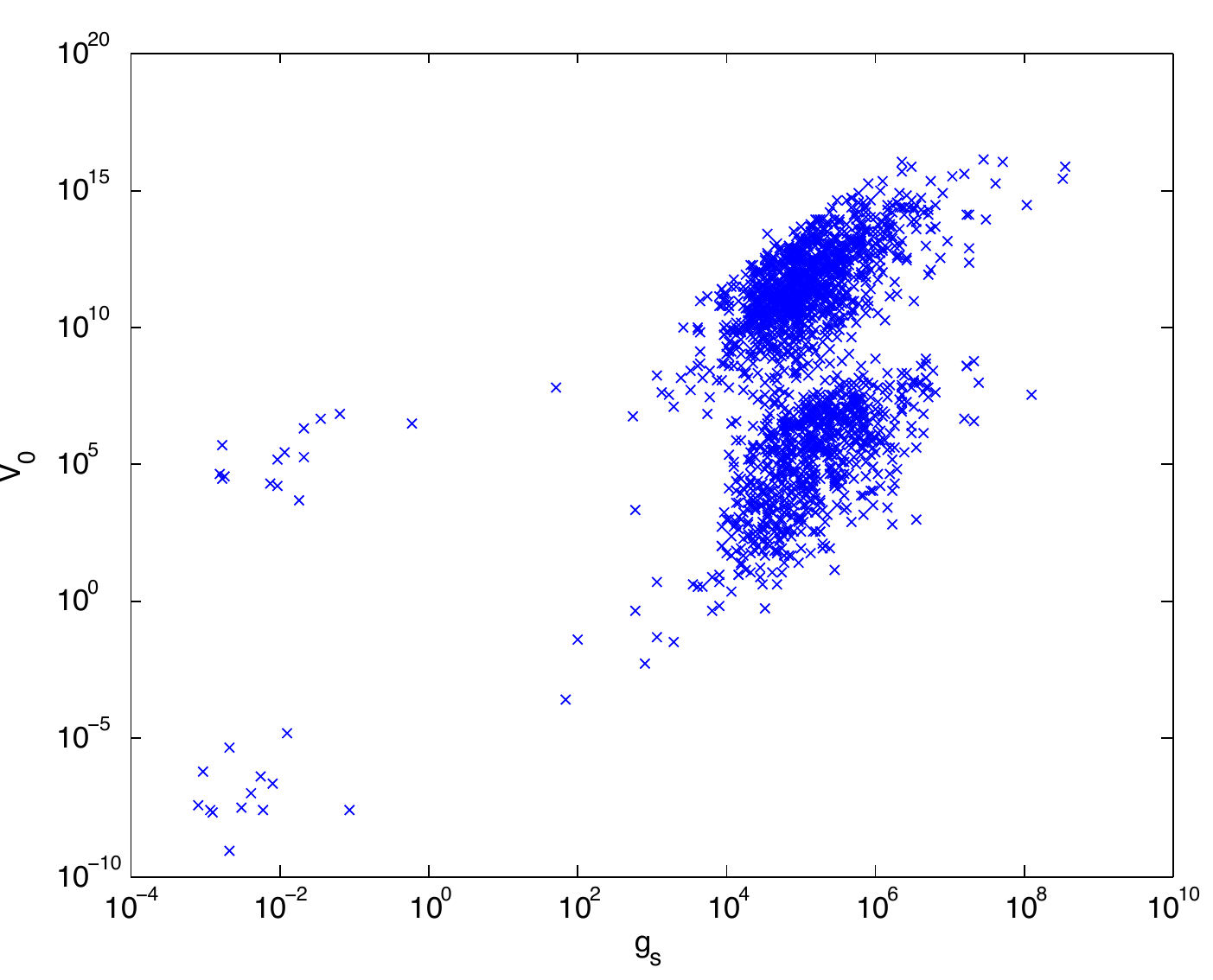} b) \includegraphics[height=5.cm]{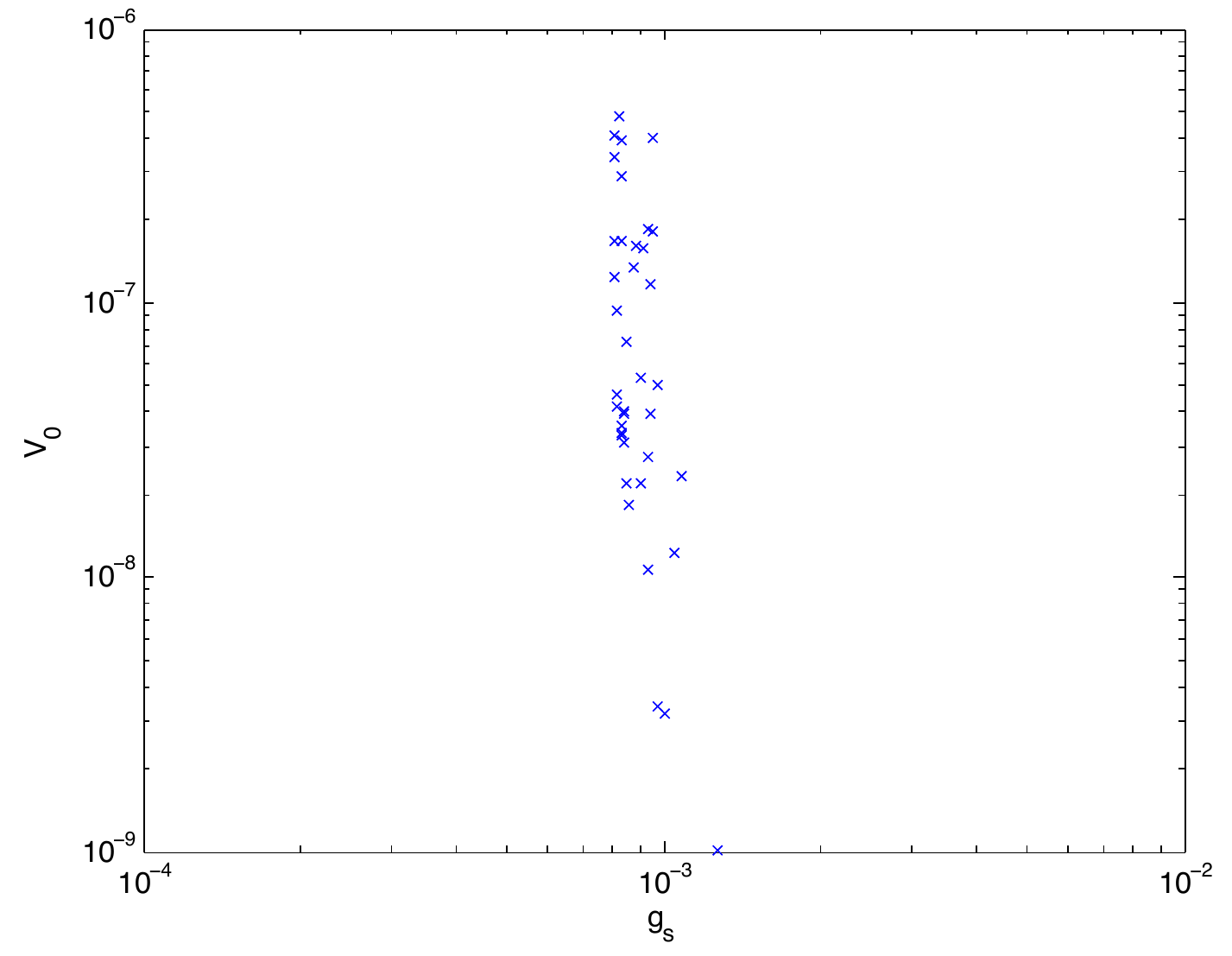}
\par\end{centering}
\caption{\label{fig:toro6} Landscape for $\tilde{N}=2$ for a) AdS and b) dS.}
\end{figure}

Here we comment on the possibility to increase the number of stable vacua by allowing the presence of odd integer fluxes. This yields to consider exotic orientifolds in our setup \cite{Frey:2002hf}. 
The presence of odd  integer fluxes implies a modification on the Tadpole constraints. For the case of  a single odd flux, the Bianchi identity is  given by $\frac{1}{2}\int F_3\wedge H_3=15$. But since all remaining fluxes are even, the identity cannot be fulfilled. For two odd fluxes (two exotic orientifolds), the Bianchi identity allows several solutions which are explored by searching the minima of the scalar potential with all the moduli stabilized. The corresponding landscape is shown in Figure \ref{fig:toro6} for which a further analysis reveals that only one case of the full-found landscape is actually stable. The moduli expectation values are given in Table \ref{tab:orientifoldvac}.\\
.\\
\begin{table}[htbp]
   \centering
     \caption{Moduli and energy vev's.}
   \begin{tabular}{@{} cccccccc @{}} 
     \hline
      $\langle{\phi_1}\rangle$    & $\langle{\psi_1}\rangle$ & $\langle{\phi_2}\rangle$ & $\langle{\psi_2}\rangle$ & $\langle{\phi_3}\rangle$ & $\langle{\psi_3}\rangle$	& $V_{min}$\\
     \hline
     0.005		& 0.0251		& -134.55		& 5461.63		& -70.085		& 1388.41		& $2.8588{\times}10^{-10}$\\    
    \hline
   \end{tabular}
   \label{tab:orientifoldvac}
\end{table}

The corresponding non-vanishing flux configuration is given by the values: $a_{00} = 2$, $a_{01} = 18$, $a_{02} = 15$,  $a_{03} = 16$, $a_{13} = 15$, $a_{22}^1 = 2$ and $a_{23} = 18$.  Notice that odd fluxes are supported in different cycles as required and that this solutions implies the presence  62 $O3^-$,  one $O3^+$ and one  $\tilde{O3}^-$, which all together contribute to a total D3-brane charge of 15. \\

The moduli masses in the stable dS vacua are given in Table \ref{mini_m_2}, from which we observe that a hierarchy is also present. Notice that masses for the $U$ and $S$ moduli are quite similar. This is a consequence of the mass term $\frac{2}{3}V_0 \mathscr{K}_{ij}$ coming from the curvature of the space-time which is much larger than the second derivative $D_i\partial_j V$. The gravitinos mass and the SUSy breaking scale are also quite high for realistic scenarios (around $10^{13}$  and $10^{21}$ GeV respectively).
A similar analysis revels that slow-roll inflation conditions are also present for small fields, with the inflaton composed by a combination of the complex fields $\Phi_2$ and $\Phi_3$. Since the features of this solution are quite similar to those without exotic orientifolds, we omit the detailed analysis. However we observe that considering the presence of odd fluxes increases the size of the field space related to stable dS vacua where slow-roll inflation is present.\\
\begin{center}
\begin{table}[h]
\caption{\label{mini_m_2} Mass  Eigenvalues for stable dS vacua for $\tilde{N}=2$.}
\centering
\begin{tabular}{c c c c c c c c}
\hline
\bf{Solution} & $\tilde{M}_{\phi_1}^2$ & $\tilde{M}_{\psi_1}^2$ & $\tilde{M}_{\phi_2}^2$ & $\tilde{M}_{\psi_2}^2$ & $\tilde{M}_{\phi_3}^2$ & $\tilde{M}_{\psi_3}^2$ \\
\hline 
\bf{1} 		& $1.341{\times}10^{-5}$ & $1.445{\times}10^{-7}$ & $1.906{\times}10^{-10}$ & $2.026{\times}10^{-10}$ & $1.906{\times}10^{-10}$ & $1.906{\times}10^{-10}$ \\
\hline 
\end{tabular}
\end{table}
\end{center}

\section{Discussion and Final Comments}
In this paper we have searched for stable de Sitter and Anti de-Sitter vacua in the context of Type IIB string compactification on a 6-dimensional isotropic torus threaded with NS-NS, R-R and non-geometric fluxes. These kind of scenarios contain all the essential ingredients for the presence of positive valued vacua. We have  performed our searching by computing the scalar potential from the corresponding superpotential which depends on all moduli at tree level.  This allows us to look for classical stable vacua within the context of a supersymmetric language.\\

We have restricted  our search to non-SUSY vacua and we have found that the single case compatible with Tadpole cancelation and Bianchi identities corresponds to the case in which SUSY is broken through all moduli, although the complex structure presents the higher $F$-term.\\

By implementing a genetic algorithm we have searched for stable vacua and  we have found that the number of stable dS and AdS vacua are of the same order.  With respect to dS vacua, we report 10 (similar) flux configurations with all constraints fulfilled in the presence of standard orientifold three-planes (i.e., all fluxes are even). In all of them, we find that the masses of the complex structure moduli are larger than the masses of the dilaton and of the K\"ahler moduli. We have also found that supersymmetry is broken through the complex structure at a very high scale (around $10^{15}$ GeV) with a gravitino mass of the order of $10^{14}$ GeV while the vev's energy  is of order $10^{10}$ GeV. We believe that this issue could be improved once we incorporate interactions with other fields via the presence of D-branes.\\

Concerning the existence of suitable conditions for inflation, we have found that in all the stable dS vacua we report, multi-field inflation driven by the axio-dilaton and the K\"ahler moduli is a viable scenario. Concretely, we have found that the inflaton is almost orthogonal to the s-goldstino direction and that the imaginary part of the dilaton and the internal volume are the real fields which contribute more significantly to the slow-roll parameter values. In particular we show that single field inflation is discarded in agreement with the bounds computed in \cite{Borghese:2012yu}.  An inflationary direction driven by the above two real fields is a viable direction according to our calculations. However, specific directions requires more detailed considerations which are beyond the scope of this work.  Finally we increase our field space by turning on  odd fluxes. As it is well known this implies the presence of exotic orientifolds. In this case, we report one single case for a stable dS vacuum with the same characteristics as the vacua found in the absence of exotic orientifolds: a high scale for SUSY breaking, high gravitino mass, a hierarchy on the moduli masses which separates the complex structure from the axio-dilaton and K\"ahler moduli. Single field inflation in the small regime is also discarded. \\

It is worth mentioning some issues which limited our study. First of all, the use of genetic algorithms makes impossible the search of Minkowski vacua since it works on a numerical approximation. Second, this method also excludes the possibility to conclude either a minimum is meta-stable or global. However we can assure that in case of having a lower minimum, it must be more than a 100 mass planck units away from the reported one. Metastable vacua close to the studied minimum are not considered by the algorithm. Finally, since the complex structure moduli decay previous to the inflaton field, it plays the role of a Polonyi field. So far we have not a solution for the cosmological moduli problem derived from this scenario. We leave this feature for further work. Finally, we should say that inflation driven by all moduli or by a single one is not discarded in a large field scenario.\\

\begin{center}
{\bf Acknowledgements}
\end{center}
We thank  David Andriot, Athanasios Chatzistavrakidis, Michele Cicoli, Rom\'an Linares, Gustavo Niz, Octavio Obregon, Saul Ramos-Sanchez and  Diederik Roest for useful suggestions  and for many interesting discussions. D. A and L. D-B are supported by a doctoral CONACyT grant. O.L.-B. was partially supported by CONACyT under contract No. 132166. M.S was partially supported by CONACyT under contracts No. 62253 and No.135023 and by DAIP 125-11.

\newpage

\appendix

\section{Scalar potential}
The explicit form of the superpotential for the stable cases reported in Table 2 is given by
\begin{equation}
 \mathscr{W}(\Phi_1,\Phi_2,\Phi_3)=a_{00}-3a_{01} \Phi_1 + 3 a_{02} \Phi_1^2-a_{03} \Phi_1^3+a_{13} \Phi_2 \Phi_1^3- 3 \Phi_3\left({a_{32}^1 \Phi_1^2+a_{33} \Phi_1^3}\right),
\end{equation}
while the explicit form for the scalar potential in terms of those fluxes is written as;\\

\begin{align*}
& V=\frac{1}{{128 \psi _1^3 \psi _2 \psi _3^3}}\big(4 a_{03}^2 \phi _1^6+4 a_{13}^2 \phi _1^6+36 a_{23}^2 \phi _3^2 \phi _1^6+12 a_{23}^2 \psi _3^2 \phi _1^6\nonumber\\
&-8 a_{03} a_{13} \phi _1^6+24 a_{03} a_{23} \phi _3 \phi _1^6-24 a_{13} a_{23} \phi _3 \phi _1^6+72 (a^2_{22}) a_{23} \phi _3^2 \phi _1^5+24 (a^2_{22}) a_{23} \psi _3^2 \phi _1^5\nonumber\\
&-24 a_{02} a_{03} \phi _1^5+24 a_{02} a_{13} \phi _1^5+24 a_{03} (a^2_{22}) \phi _3 \phi _1^5-24 a_{13} (a^2_{22}) \phi _3 \phi _1^5-72 a_{02} a_{23} \phi _3 \phi _1^5+36 a_{02}^2 \phi _1^4\nonumber\\
&+36 (a^2_{22})^2 \phi _3^2 \phi _1^4+12 a_{03}^2 \psi _1^2 \phi _1^4+12 a_{13}^2 \psi _1^2 \phi _1^4+108 a_{23}^2 \phi _3^2 \psi _1^2 \phi _1^4-24 a_{03} a_{13} \psi _1^2 \phi _1^4\nonumber\\
&+72 a_{03} a_{23} \phi _3 \psi _1^2 \phi _1^4-72 a_{13} a_{23} \phi _3 \psi _1^2 \phi _1^4+36 (a^2_{22})^2 \psi _3^2 \phi _1^4+36 a_{23}^2 \psi _1^2 \psi _3^2 \phi _1^4\nonumber\\
&+24 a_{01} a_{03} \phi _1^4-24 a_{01} a_{13} \phi _1^4-72 a_{02} (a^2_{22}) \phi _3 \phi _1^4+72 a_{01} a_{23} \phi _3 \phi _1^4+144 (a^2_{22}) a_{23} \phi _3^2 \psi _1^2 \phi _1^3\nonumber\\
&-48 a_{02} a_{03} \psi _1^2 \phi _1^3+48 a_{02} a_{13} \psi _1^2 \phi _1^3+48 a_{03} (a^2_{22}) \phi _3 \psi _1^2 \phi _1^3-48 a_{13} (a^2_{22}) \phi _3 \psi _1^2 \phi _1^3\nonumber\\
&-144 a_{02} a_{23} \phi _3 \psi _1^2 \phi _1^3+48 (a^2_{22}) a_{23} \psi _1^2 \psi _3^2 \phi _1^3-72 a_{01} a_{02} \phi _1^3-8 a_{00} a_{03} \phi _1^3+8 a_{00} a_{13} \phi _1^3\nonumber\\
&+72 a_{01} (a^2_{22}) \phi _3 \phi _1^3-24 a_{00} a_{23} \phi _3 \phi _1^3+12 a_{03}^2 \psi _1^4 \phi _1^2+12 a_{13}^2 \psi _1^4 \phi _1^2+108 a_{23}^2 \phi _3^2 \psi _1^4 \phi _1^2\nonumber\\
&-24 a_{03} a_{13} \psi _1^4 \phi _1^2+72 a_{03} a_{23} \phi _3 \psi _1^4 \phi _1^2-72 a_{13} a_{23} \phi _3 \psi _1^4 \phi _1^2+36 a_{01}^2 \phi _1^2+48 a_{02}^2 \psi _1^2 \phi _1^2\nonumber\\
&+48 (a^2_{22})^2 \phi _3^2 \psi _1^2 \phi _1^2+24 a_{01} a_{03} \psi _1^2 \phi _1^2-24 a_{01} a_{13} \psi _1^2 \phi _1^2-96 a_{02} (a^2_{22}) \phi _3 \psi _1^2 \phi _1^2\nonumber\\
&+72 a_{01} a_{23} \phi _3 \psi _1^2 \phi _1^2+36 a_{23}^2 \psi _1^4 \psi _3^2 \phi _1^2+48 (a^2_{22})^2 \psi _1^2 \psi _3^2 \phi _1^2+24 a_{00} a_{02} \phi _1^2\nonumber\\
&-24 a_{00} (a^2_{22}) \phi _3 \phi _1^2+72 (a^2_{22}) a_{23} \phi _3^2 \psi _1^4 \phi _1-24 a_{02} a_{03} \psi _1^4 \phi _1+24 a_{02} a_{13} \psi _1^4 \phi _1\nonumber\\
&+24 a_{03} (a^2_{22}) \phi _3 \psi _1^4 \phi _1-24 a_{13} (a^2_{22}) \phi _3 \psi _1^4 \phi _1-72 a_{02} a_{23} \phi _3 \psi _1^4 \phi _1-48 a_{01} a_{02} \psi _1^2 \phi _1\nonumber\\
&+48 a_{01} (a^2_{22}) \phi _3 \psi _1^2 \phi _1+24 (a^2_{22}) a_{23} \psi _1^4 \psi _3^2 \phi _1-24 a_{00} a_{01} \phi _1+4 a_{03}^2 \psi _1^6+4 a_{13}^2 \psi _1^6\nonumber\\
&+36 a_{23}^2 \phi _3^2 \psi _1^6-8 a_{03} a_{13} \psi _1^6+24 a_{03} a_{23} \phi _3 \psi _1^6-24 a_{13} a_{23} \phi _3 \psi _1^6+12 a_{02}^2 \psi _1^4\nonumber\\
&+12 (a^2_{22})^2 \phi _3^2 \psi _1^4-24 a_{02} (a^2_{22}) \phi _3 \psi _1^4+4 a_{00}^2+12 a_{01}^2 \psi _1^2+12 a_{23}^2 \psi _1^6 \psi _3^2+12 (a^2_{22})^2 \psi _1^4 \psi _3^2\nonumber\\
&+24 a_{01} (a^2_{22}) \psi _1^3 \psi _3+24 a_{00} a_{23} \psi _1^3 \psi _3\big).
\end{align*}

For completeness we give the expression for the scalar potential with all fluxes turned on:
{\tiny
\begin{align*}
&V=\frac{1}{128 \psi _1^3 \psi _2 \psi _3^3}\big(4 a_{03}^2 \phi _1^6+4 a_{13}^2 \phi _1^6+36 a_{23}^2 \phi _3^2 \phi _1^6+12 a_{23}^2 \psi _3^2 \phi _1^6-8 a_{03} a_{13} \phi _1^6+24 a_{03} a_{23} \phi _3 \phi _1^6-24 a_{13} a_{23} \phi _3 \phi _1^6\nonumber\\
&+144 (a^2_{22}) a_{23} \phi _3^2 \phi _1^5+72 (a^1_{22}) a_{23} \phi _3^2 \phi _1^5+48 (a^2_{22}) a_{23} \psi _3^2 \phi _1^5+24 (a^1_{22}) a_{23} \psi _3^2 \phi _1^5-24 a_{02} a_{03} \phi _1^5+24 a_{03} a_{12} \phi _1^5+24 a_{02} a_{13} \phi _1^5\nonumber\\
&-24 a_{12} a_{13} \phi _1^5+48 a_{03} (a^2_{22}) \phi _3 \phi _1^5-48 a_{13} (a^2_{22}) \phi _3 \phi _1^5+24 a_{03} (a^1_{22}) \phi _3 \phi _1^5-24 a_{13} (a^1_{22}) \phi _3 \phi _1^5-72 a_{02} a_{23} \phi _3 \phi _1^5+72 a_{12} a_{23} \phi _3 \phi _1^5\nonumber\\
&+36 a_{02}^2 \phi _1^4+36 a_{12}^2 \phi _1^4+144 (a^2_{22})^2 \phi _3^2 \phi _1^4+36 (a^1_{22})^2 \phi _3^2 \phi _1^4+144 (a^2_{22}) (a^1_{22}) \phi _3^2 \phi _1^4-144 (a^2_{22}) a_{23} \phi _3^2 \phi _1^4-72 (a^1_{22}) a_{23} \phi _3^2 \phi _1^4+12 a_{03}^2 \psi _1^2 \phi _1^4\nonumber\\
&+12 a_{13}^2 \psi _1^2 \phi _1^4+108 a_{23}^2 \phi _3^2 \psi _1^2 \phi _1^4-24 a_{03} a_{13} \psi _1^2 \phi _1^4+72 a_{03} a_{23} \phi _3 \psi _1^2 \phi _1^4-72 a_{13} a_{23} \phi _3 \psi _1^2 \phi _1^4+72 (a^2_{22})^2 \psi _3^2 \phi _1^4\nonumber\\
&+36 (a^1_{22})^2 \psi _3^2 \phi _1^4+36 a_{23}^2 \psi _1^2 \psi _3^2 \phi _1^4-48 (a^2_{22}) a_{23} \psi _3^2 \phi _1^4-24 (a^1_{22}) a_{23} \psi _3^2 \phi _1^4+24 a_{01} a_{03} \phi _1^4-24 a_{03} a_{11} \phi _1^4-72 a_{02} a_{12} \phi _1^4\nonumber\\
&-24 a_{01} a_{13} \phi _1^4+24 a_{11} a_{13} \phi _1^4-48 a_{03} (a^2_{22}) \phi _3 \phi _1^4+48 a_{13} (a^2_{22}) \phi _3 \phi _1^4-24 a_{03} (a^1_{22}) \phi _3 \phi _1^4+24 a_{13} (a^1_{22}) \phi _3 \phi _1^4-144 a_{02} (a^2_{22}) \phi _3 \phi _1^4\nonumber\\
&+144 a_{12} (a^2_{22}) \phi _3 \phi _1^4-72 a_{02} (a^1_{22}) \phi _3 \phi _1^4+72 a_{12} (a^1_{22}) \phi _3 \phi _1^4+72 a_{01} a_{23} \phi _3 \phi _1^4-72 a_{11} a_{23} \phi _3 \phi _1^4-288 (a^2_{22}) (a^2_{22}) \phi _3^2 \phi _1^3\nonumber\\
&-144 (a^1_{22}) (a^2_{22}) \phi _3^2 \phi _1^3-144 (a^2_{22}) (a^1_{22}) \phi _3^2 \phi _1^3-72 (a^1_{22}) (a^1_{22}) \phi _3^2 \phi _1^3-72 a_{20} a_{23} \phi _3^2 \phi _1^3+288 (a^2_{22}) a_{23} \phi _3^2 \psi _1^2 \phi _1^3+144 (a^1_{22}) a_{23} \phi _3^2 \psi _1^2 \phi _1^3\nonumber\\
&-48 a_{02} a_{03} \psi _1^2 \phi _1^3+48 a_{03} a_{12} \psi _1^2 \phi _1^3+48 a_{02} a_{13} \psi _1^2 \phi _1^3-48 a_{12} a_{13} \psi _1^2 \phi _1^3+96 a_{03} (a^2_{22}) \phi _3 \psi _1^2 \phi _1^3-96 a_{13} (a^2_{22}) \phi _3 \psi _1^2 \phi _1^3\nonumber\\
&+48 a_{03} (a^1_{22}) \phi _3 \psi _1^2 \phi _1^3-48 a_{13} (a^1_{22}) \phi _3 \psi _1^2 \phi _1^3-144 a_{02} a_{23} \phi _3 \psi _1^2 \phi _1^3+144 a_{12} a_{23} \phi _3 \psi _1^2 \phi _1^3+96 (a^2_{22}) a_{23} \psi _1^2 \psi _3^2 \phi _1^3\nonumber\\
&+48 (a^1_{22}) a_{23} \psi _1^2 \psi _3^2 \phi _1^3-144 (a^2_{22}) (a^2_{22}) \psi _3^2 \phi _1^3-72 (a^1_{22}) (a^1_{22}) \psi _3^2 \phi _1^3-24 a_{20} a_{23} \psi _3^2 \phi _1^3-72 a_{01} a_{02} \phi _1^3-8 a_{00} a_{03} \phi _1^3+72 a_{02} a_{11} \phi _1^3\nonumber\\
&+72 a_{01} a_{12} \phi _1^3-72 a_{11} a_{12} \phi _1^3+8 a_{00} a_{13} \phi _1^3+8 a_{03} a_{10} \phi _2 \phi _1^3-8 a_{10} a_{13} \phi _2 \phi _1^3-24 a_{03} a_{20} \phi _3 \phi _1^3+24 a_{13} a_{20} \phi _3 \phi _1^3\nonumber\\
&+144 a_{02} (a^2_{22}) \phi _3 \phi _1^3-144 a_{12} (a^2_{22}) \phi _3 \phi _1^3+72 a_{02} (a^1_{22}) \phi _3 \phi _1^3-72 a_{12} (a^1_{22}) \phi _3 \phi _1^3+144 a_{01} (a^2_{22}) \phi _3 \phi _1^3-144 a_{11} (a^2_{22}) \phi _3 \phi _1^3\nonumber\\
&+72 a_{01} (a^1_{22}) \phi _3 \phi _1^3-72 a_{11} (a^1_{22}) \phi _3 \phi _1^3-24 a_{00} a_{23} \phi _3 \phi _1^3+24 a_{10} a_{23} \phi _2 \phi _3 \phi _1^3+12 a_{03}^2 \psi _1^4 \phi _1^2+12 a_{13}^2 \psi _1^4 \phi _1^2+108 a_{23}^2 \phi _3^2 \psi _1^4 \phi _1^2\nonumber\\
&-24 a_{03} a_{13} \psi _1^4 \phi _1^2+72 a_{03} a_{23} \phi _3 \psi _1^4 \phi _1^2-72 a_{13} a_{23} \phi _3 \psi _1^4 \phi _1^2+36 a_{01}^2 \phi _1^2+36 a_{11}^2 \phi _1^2+144 (a^2_{22})^2 \phi _3^2 \phi _1^2+36 (a^1_{22})^2 \phi _3^2 \phi _1^2\nonumber\\
&+144 (a^2_{22}) (a^1_{22}) \phi _3^2 \phi _1^2-144 a_{20} (a^2_{22}) \phi _3^2 \phi _1^2-72 a_{20} (a^1_{22}) \phi _3^2 \phi _1^2+48 a_{02}^2 \psi _1^2 \phi _1^2+48 a_{12}^2 \psi _1^2 \phi _1^2+192 (a^2_{22})^2 \phi _3^2 \psi _1^2 \phi _1^2\nonumber\\
&+48 (a^1_{22})^2 \phi _3^2 \psi _1^2 \phi _1^2+192 (a^2_{22}) (a^1_{22}) \phi _3^2 \psi _1^2 \phi _1^2-144 (a^2_{22}) a_{23} \phi _3^2 \psi _1^2 \phi _1^2-72 (a^1_{22}) a_{23} \phi _3^2 \psi _1^2 \phi _1^2+24 a_{01} a_{03} \psi _1^2 \phi _1^2\nonumber\\
&-24 a_{03} a_{11} \psi _1^2 \phi _1^2-96 a_{02} a_{12} \psi _1^2 \phi _1^2-24 a_{01} a_{13} \psi _1^2 \phi _1^2+24 a_{11} a_{13} \psi _1^2 \phi _1^2-48 a_{03} (a^2_{22}) \phi _3 \psi _1^2 \phi _1^2+48 a_{13} (a^2_{22}) \phi _3 \psi _1^2 \phi _1^2\nonumber\\
&-24 a_{03} (a^1_{22}) \phi _3 \psi _1^2 \phi _1^2+24 a_{13} (a^1_{22}) \phi _3 \psi _1^2 \phi _1^2-192 a_{02} (a^2_{22}) \phi _3 \psi _1^2 \phi _1^2+192 a_{12} (a^2_{22}) \phi _3 \psi _1^2 \phi _1^2-96 a_{02} (a^1_{22}) \phi _3 \psi _1^2 \phi _1^2\nonumber\\
&+96 a_{12} (a^1_{22}) \phi _3 \psi _1^2 \phi _1^2+72 a_{01} a_{23} \phi _3 \psi _1^2 \phi _1^2-72 a_{11} a_{23} \phi _3 \psi _1^2 \phi _1^2+36 a_{23}^2 \psi _1^4 \psi _3^2 \phi _1^2+72 (a^2_{22})^2 \psi _3^2 \phi _1^2+36 (a^1_{22})^2 \psi _3^2 \phi _1^2\nonumber\\
&+48 (a^2_{22})^2 \psi _1^2 \psi _3^2 \phi _1^2+48 (a^1_{22})^2 \psi _1^2 \psi _3^2 \phi _1^2-96 (a^2_{22}) (a^1_{22}) \psi _1^2 \psi _3^2 \phi _1^2-144 (a^2_{22}) a_{23} \psi _1^2 \psi _3^2 \phi _1^2-72 (a^1_{22}) a_{23} \psi _1^2 \psi _3^2 \phi _1^2\nonumber\\
&-48 a_{20} (a^2_{22}) \psi _3^2 \phi _1^2-24 a_{20} (a^1_{22}) \psi _3^2 \phi _1^2+24 a_{00} a_{02} \phi _1^2-72 a_{01} a_{11} \phi _1^2-24 a_{00} a_{12} \phi _1^2-24 a_{02} a_{10} \phi _2 \phi _1^2+24 a_{10} a_{12} \phi _2 \phi _1^2\nonumber\\
&+72 a_{02} a_{20} \phi _3 \phi _1^2-72 a_{12} a_{20} \phi _3 \phi _1^2-144 a_{01} (a^2_{22}) \phi _3 \phi _1^2+144 a_{11} (a^2_{22}) \phi _3 \phi _1^2-72 a_{01} (a^1_{22}) \phi _3 \phi _1^2+72 a_{11} (a^1_{22}) \phi _3 \phi _1^2-48 a_{00} (a^2_{22}) \phi _3 \phi _1^2\nonumber\\
&-24 a_{00} (a^1_{22}) \phi _3 \phi _1^2+48 a_{10} (a^2_{22}) \phi _2 \phi _3 \phi _1^2+24 a_{10} (a^1_{22}) \phi _2 \phi _3 \phi _1^2+144 (a^2_{22}) a_{23} \phi _3^2 \psi _1^4 \phi _1+72 (a^1_{22}) a_{23} \phi _3^2 \psi _1^4 \phi _1-24 a_{02} a_{03} \psi _1^4 \phi _1\nonumber\\
&+24 a_{03} a_{12} \psi _1^4 \phi _1+24 a_{02} a_{13} \psi _1^4 \phi _1-24 a_{12} a_{13} \psi _1^4 \phi _1+48 a_{03} (a^2_{22}) \phi _3 \psi _1^4 \phi _1-48 a_{13} (a^2_{22}) \phi _3 \psi _1^4 \phi _1+24 a_{03} (a^1_{22}) \phi _3 \psi _1^4 \phi _1\nonumber\\
&-24 a_{13} (a^1_{22}) \phi _3 \psi _1^4 \phi _1-72 a_{02} a_{23} \phi _3 \psi _1^4 \phi _1+72 a_{12} a_{23} \phi _3 \psi _1^4 \phi _1+144 a_{20} (a^2_{22}) \phi _3^2 \phi _1+72 a_{20} (a^1_{22}) \phi _3^2 \phi _1-192 (a^2_{22}) (a^2_{22}) \phi _3^2 \psi _1^2 \phi _1\nonumber\\
&-96 (a^1_{22}) (a^2_{22}) \phi _3^2 \psi _1^2 \phi _1-96 (a^2_{22}) (a^1_{22}) \phi _3^2 \psi _1^2 \phi _1-48 (a^1_{22}) (a^1_{22}) \phi _3^2 \psi _1^2 \phi _1-48 a_{01} a_{02} \psi _1^2 \phi _1+48 a_{02} a_{11} \psi _1^2 \phi _1+48 a_{01} a_{12} \psi _1^2 \phi _1\nonumber\\
&-48 a_{11} a_{12} \psi _1^2 \phi _1+96 a_{02} (a^2_{22}) \phi _3 \psi _1^2 \phi _1-96 a_{12} (a^2_{22}) \phi _3 \psi _1^2 \phi _1+48 a_{02} (a^1_{22}) \phi _3 \psi _1^2 \phi _1-48 a_{12} (a^1_{22}) \phi _3 \psi _1^2 \phi _1+96 a_{01} (a^2_{22}) \phi _3 \psi _1^2 \phi _1\nonumber\\
&-96 a_{11} (a^2_{22}) \phi _3 \psi _1^2 \phi _1+48 a_{01} (a^1_{22}) \phi _3 \psi _1^2 \phi _1-48 a_{11} (a^1_{22}) \phi _3 \psi _1^2 \phi _1+48 (a^2_{22}) a_{23} \psi _1^4 \psi _3^2 \phi _1+24 (a^1_{22}) a_{23} \psi _1^4 \psi _3^2 \phi _1\nonumber\\
&-48 (a^2_{22}) (a^2_{22}) \psi _1^2 \psi _3^2 \phi _1+48 (a^1_{22}) (a^2_{22}) \psi _1^2 \psi _3^2 \phi _1+48 (a^2_{22}) (a^1_{22}) \psi _1^2 \psi _3^2 \phi _1\nonumber\\
&-48 (a^1_{22}) (a^1_{22}) \psi _1^2 \psi _3^2 \phi _1-144 a_{20} a_{23} \psi _1^2 \psi _3^2 \phi _1+48 a_{20} (a^2_{22}) \psi _3^2 \phi _1+24 a_{20} (a^1_{22}) \psi _3^2 \phi _1-24 a_{00} a_{01} \phi _1+24 a_{00} a_{11} \phi _1+24 a_{01} a_{10} \phi _2 \phi _1\nonumber\\
&-24 a_{10} a_{11} \phi _2 \phi _1-72 a_{01} a_{20} \phi _3 \phi _1+72 a_{11} a_{20} \phi _3 \phi _1+48 a_{00} (a^2_{22}) \phi _3 \phi _1+24 a_{00} (a^1_{22}) \phi _3 \phi _1-48 a_{10} (a^2_{22}) \phi _2 \phi _3 \phi _1-24 a_{10} (a^1_{22}) \phi _2 \phi _3 \phi _1\nonumber\\
&+72 a_{10} a_{23} \psi _1^2 \psi _2 \psi _3 \phi _1+4 a_{03}^2 \psi _1^6+4 a_{13}^2 \psi _1^6+36 a_{23}^2 \phi _3^2 \psi _1^6-8 a_{03} a_{13} \psi _1^6+24 a_{03} a_{23} \phi _3 \psi _1^6-24 a_{13} a_{23} \phi _3 \psi _1^6+12 a_{02}^2 \psi _1^4\nonumber\\
&+12 a_{12}^2 \psi _1^4+48 (a^2_{22})^2 \phi _3^2 \psi _1^4+12 (a^1_{22})^2 \phi _3^2 \psi _1^4+48 (a^2_{22}) (a^1_{22}) \phi _3^2 \psi _1^4-24 a_{02} a_{12} \psi _1^4-48 a_{02} (a^2_{22}) \phi _3 \psi _1^4+48 a_{12} (a^2_{22}) \phi _3 \psi _1^4\nonumber\\
&-24 a_{02} (a^1_{22}) \phi _3 \psi _1^4+24 a_{12} (a^1_{22}) \phi _3 \psi _1^4+4 a_{00}^2+4 a_{10}^2 \phi _2^2+36 a_{20}^2 \phi _3^2+12 a_{01}^2 \psi _1^2+12 a_{11}^2 \psi _1^2+48 (a^2_{22})^2 \phi _3^2 \psi _1^2+12 (a^1_{22})^2 \phi _3^2 \psi _1^2\nonumber\\
&+48 (a^2_{22}) (a^1_{22}) \phi _3^2 \psi _1^2-24 a_{01} a_{11} \psi _1^2-48 a_{01} (a^2_{22}) \phi _3 \psi _1^2+48 a_{11} (a^2_{22}) \phi _3 \psi _1^2-24 a_{01} (a^1_{22}) \phi _3 \psi _1^2+24 a_{11} (a^1_{22}) \phi _3 \psi _1^2+4 a_{10}^2 \psi _2^2\nonumber\\
&+12 a_{23}^2 \psi _1^6 \psi _3^2-24 (a^2_{22})^2 \psi _1^4 \psi _3^2+12 (a^1_{22})^2 \psi _1^4 \psi _3^2-96 (a^2_{22}) (a^1_{22}) \psi _1^4 \psi _3^2-96 (a^2_{22}) a_{23} \psi _1^4 \psi _3^2-48 (a^1_{22}) a_{23} \psi _1^4 \psi _3^2+12 a_{20}^2 \psi _3^2\nonumber\\
&-24 (a^2_{22})^2 \psi _1^2 \psi _3^2+12 (a^1_{22})^2 \psi _1^2 \psi _3^2-96 (a^2_{22}) (a^1_{22}) \psi _1^2 \psi _3^2-96 a_{20} (a^2_{22}) \psi _1^2 \psi _3^2-48 a_{20} (a^1_{22}) \psi _1^2 \psi _3^2-8 a_{00} a_{10} \phi _2+24 a_{00} a_{20} \phi _3\nonumber\\
&-24 a_{10} a_{20} \phi _2 \phi _3+8 a_{03} a_{10} \psi _1^3 \psi _2-8 a_{10} a_{13} \psi _1^3 \psi _2+24 a_{10} a_{23} \phi _3 \psi _1^3 \psi _2-24 a_{03} a_{20} \psi _1^3 \psi _3+24 a_{13} a_{20} \psi _1^3 \psi _3-48 a_{02} (a^2_{22}) \psi _1^3 \psi _3\nonumber\\
&+48 a_{12} (a^2_{22}) \psi _1^3 \psi _3-24 a_{02} (a^1_{22}) \psi _1^3 \psi _3+24 a_{12} (a^1_{22}) \psi _1^3 \psi _3+48 a_{01} (a^2_{22}) \psi _1^3 \psi _3-48 a_{11} (a^2_{22}) \psi _1^3 \psi _3+24 a_{01} (a^1_{22}) \psi _1^3 \psi _3-24 a_{11} (a^1_{22}) \psi _1^3 \psi _3\nonumber\\
&+24 a_{00} a_{23} \psi _1^3 \psi _3-24 a_{10} a_{23} \phi _2 \psi _1^3 \psi _3+48 a_{10} (a^2_{22}) \psi _1^2 \psi _2 \psi _3+24 a_{10} (a^1_{22}) \psi _1^2 \psi _2 \psi _3\big)
\end{align*}
}

\begin{figure}[hm]
\begin{centering}
a) \includegraphics[height=5.cm]{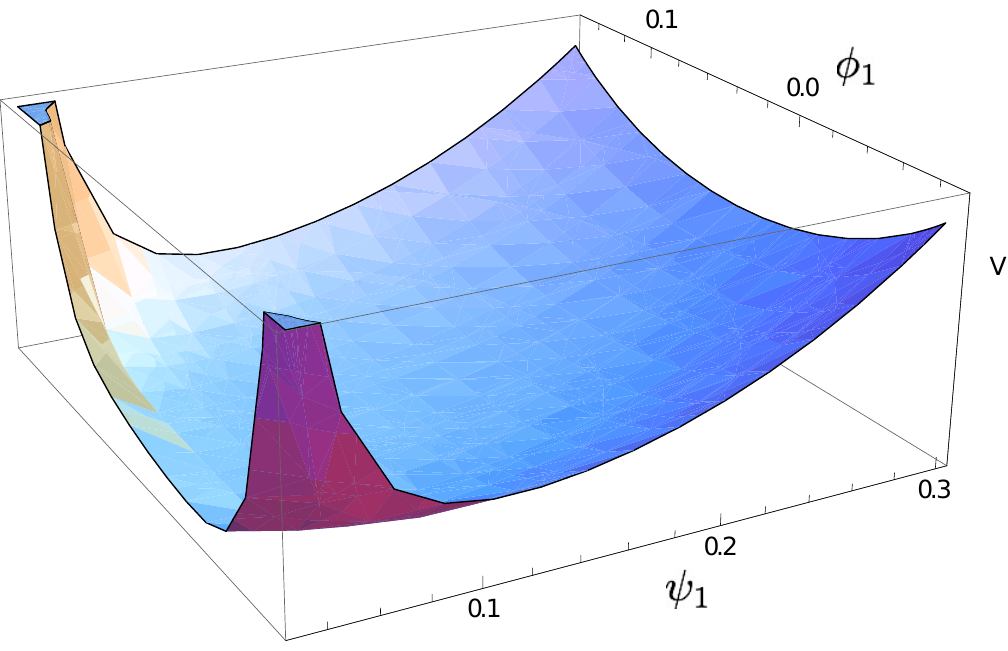} b) \includegraphics[height=5.cm]{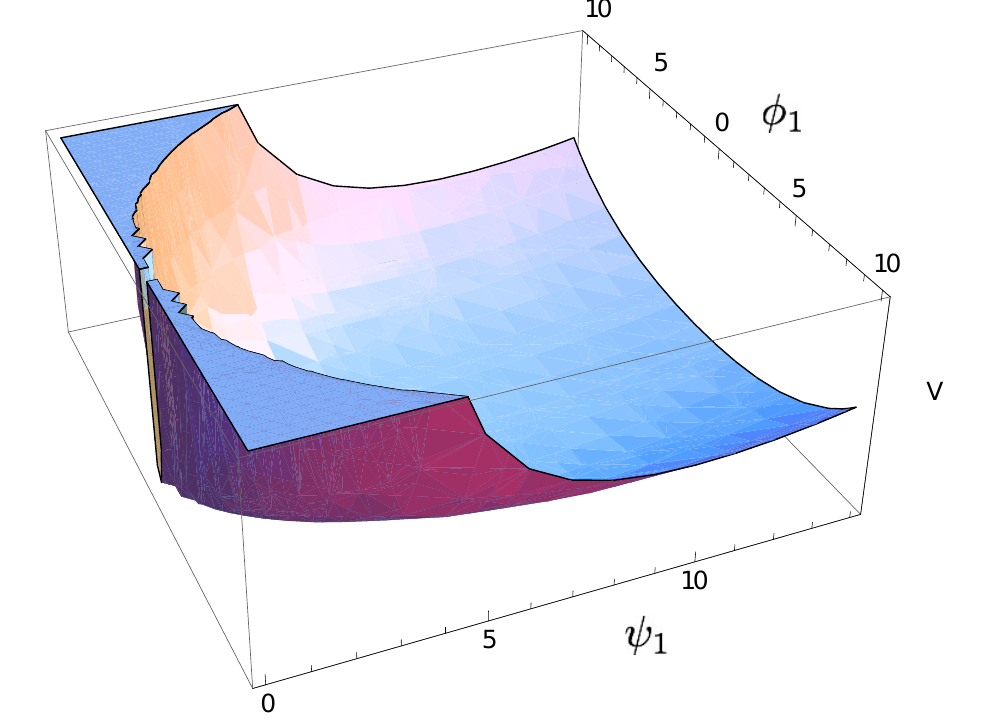}\\
c) \includegraphics[height=5.cm]{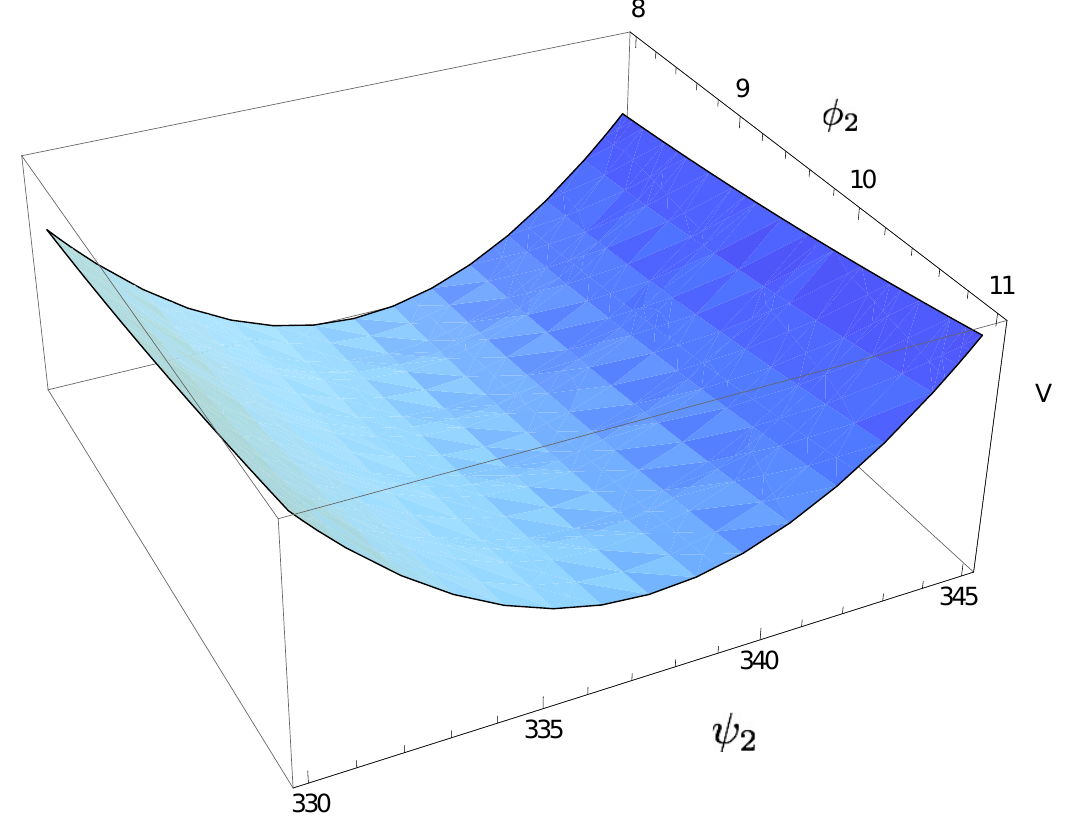} d) \includegraphics[height=5.cm]{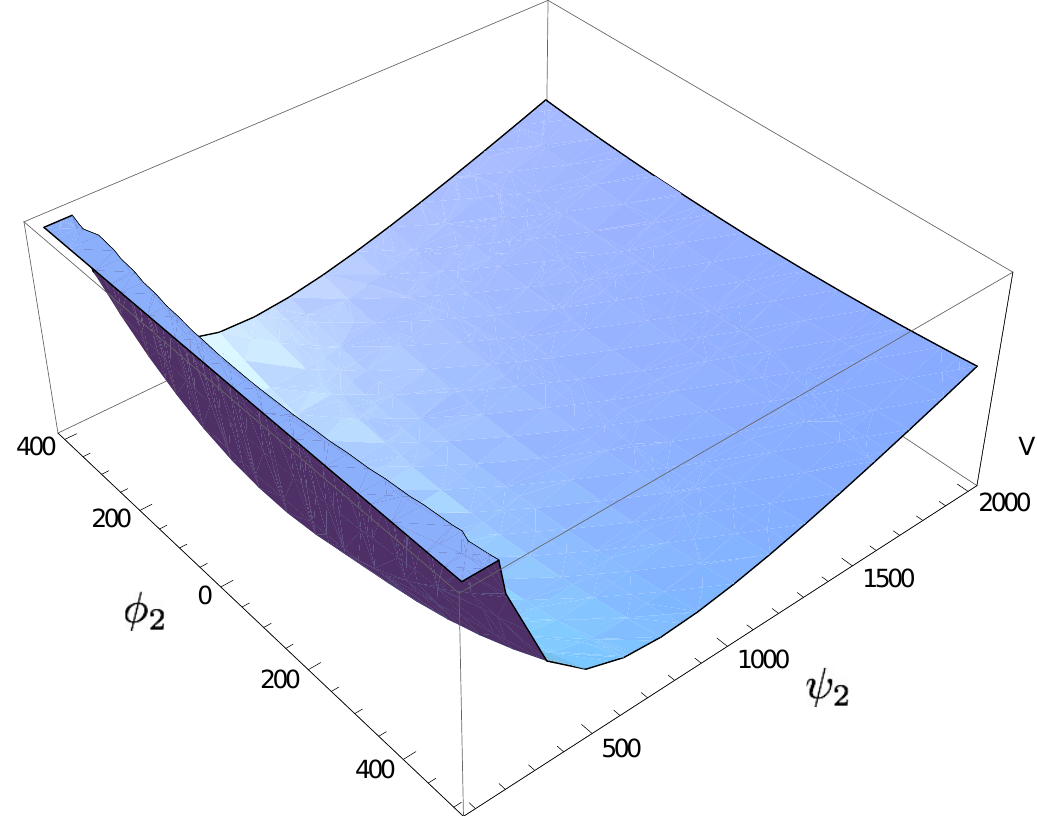}\\
e) \includegraphics[height=5.cm]{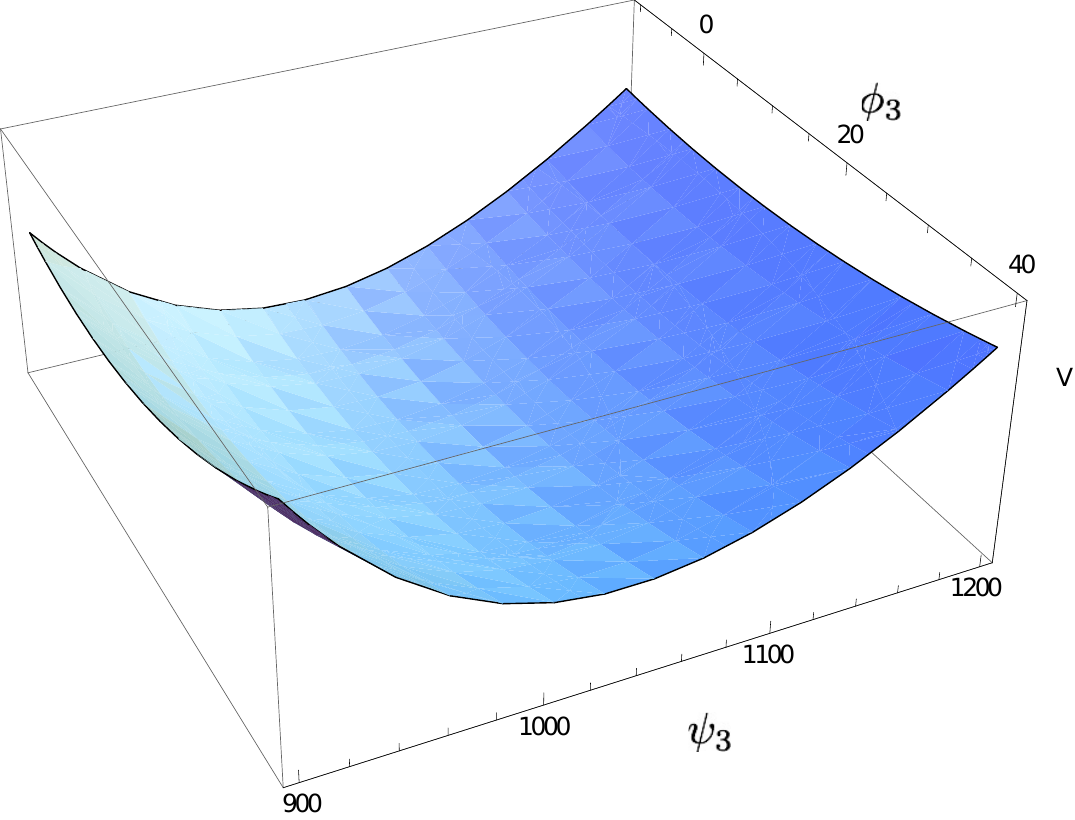} f) \includegraphics[height=5.cm]{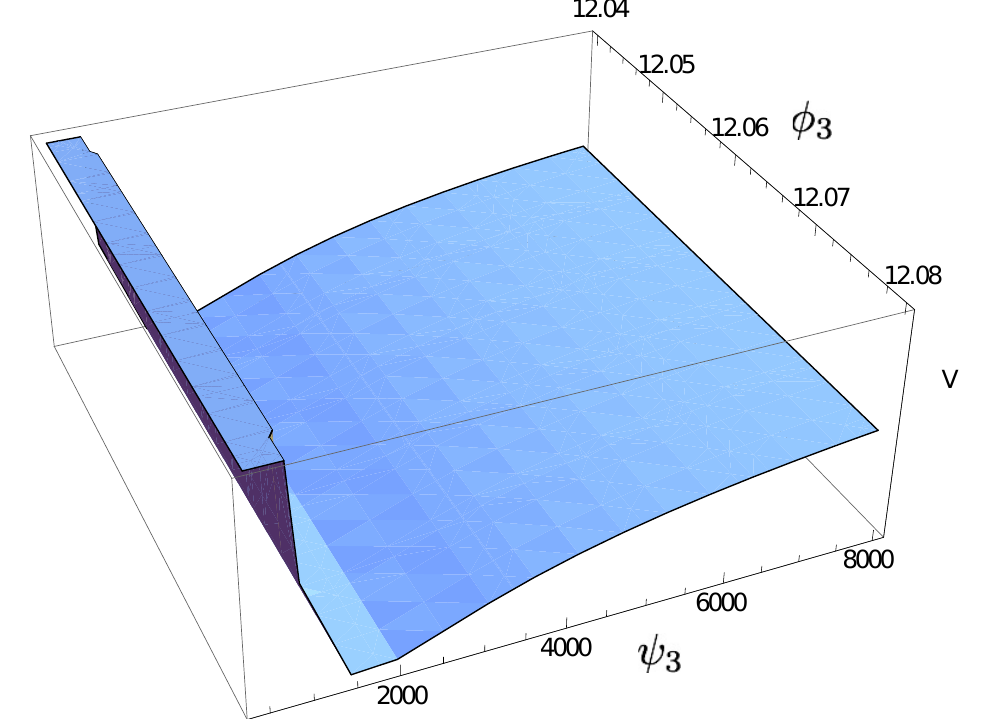}
\par\end{centering}
\caption{\label{fig:minimiza}  a) Complex Structure, c) Dilaton, e) K\"ahler modulus  for $\tilde{N}=0$, b) Complex Structure d) Dilaton   f) K\"ahler modulus for $\tilde{N}=2$.}
\end{figure}

In Figure (\ref{fig:minimiza}) we show the plots concerning the minima of the potential for Case 1 in Table 2 with a positive valued minima in the absence of exotic orientifold planes ($\tilde{N}=0$ ) and for the single case reported in the presence of two exotic orientifold planes ($\tilde{N}=2$). Observe that in both cases,  the complex structure directions of the potential are not as flat as along other directions. 

\section{Notation}
\label{ap:notation}
In this section we present for completeness the notation we follow in this work and its relation with the one commonly used in the literature. As usual the holomorphic 3-form can be written as;\\
\begin{equation}
\Omega=(dx^1+\tau{dy^1}){\wedge}(dx^2+\tau{dy^2}){\wedge}(dx^3+\tau{dy^3})
\end{equation}
where $y^i=x^{i+3}$. The orientifold  $O3$ involution acts as $\sigma{(x^i)}=-x^i$ for $i=1,...,6$. The 3-form with one leg on each torus are;\\
\begin{equation}
\begin{matrix}
\alpha_0=dx^1{\wedge}dx^2{\wedge}dx^3 & \beta_0=dy^1{\wedge}dy^2{\wedge}dy^3\\
\alpha_1=dx^1{\wedge}dy^2{\wedge}dy^3 & \beta_1=dy^1{\wedge}dx^2{\wedge}dx^3\\
\alpha_2=dy^1{\wedge}dx^2{\wedge}dy^3 & \beta_2=dx^1{\wedge}dy^2{\wedge}dx^3\\
\alpha_3=dy^1{\wedge}dy^2{\wedge}dx^3 & \beta_3=dx^1{\wedge}dx^2{\wedge}dy^3,
\end{matrix}
\end{equation}
with the normalization  $\int_{T^6}\alpha_{I}\wedge\beta_{J}=\delta_{IJ}$. So, the holomorphic 3-form is rewritten as\\
\begin{equation}
\Omega=\alpha_0+\tau^{3}\beta_{0}+\sum_{i=1}^{3}{(\tau\beta_i+\tau^2\alpha_i)},
\end{equation}
and the closed 2-forms and their 4-forms duals are\\
\begin{equation}
\begin{matrix}
\omega_i=-dx^{i}{\wedge}dy^{i} & \tilde{\omega}_k=dx^i{\wedge}dy^i{\wedge}dx^j{\wedge}dy^j,
\end{matrix}
\end{equation}
where  $\int_{T^6}\omega_{I}\wedge\tilde{\omega}_{J}=\delta_{IJ}$. The NS-NS flux  $H_3$ is written as;\\
\begin{equation}
H_3=a_{10}\beta_0+a_{13}\alpha_0+\sum_{i=0}^{3}{(a_{11}\alpha_i+a_{12}\beta_i)},
\end{equation}
while the RR 3-form flux reads\
\begin{equation}
F_3=a_{00}\beta_0+a_{03}\alpha_0+\sum_{i=0}^{3}{(a_{01}\alpha_i+a_{02}\beta_i)}.
\end{equation}
Finally,  the Kahler moduli can be written as \\
\begin{equation}
U=C_4 +\frac{1}{2}e^{-\phi}~J\wedge J=\sum_iU_i~\widetilde{\omega}^i,
\end{equation}
where $J$ is the usual K\"ahler (1,1)-form.\\

The integrated non-geometric fluxes $Q$ are given by the numbers $a_{2j}$ as shown in Table  \ref{coeff}.\\

\begin{center}
\begin{table}[h]
\caption{\label{coeff} Fluxes in the duality invariant superpotential.}
\centering
\begin{tabular}{@{}*{7}{l}}
\hline
Term&IIB Flux integer& Integer flux\\
\hline
$1$&$\bar{F}_{ijk}$&$a_{00}$\\
$\Phi_1$ & $\bar{F}_{ij\gamma}$ & $a_{01}$\\
$\Phi_1^2$ & $\bar{F}_{i{\beta}\gamma}$ & $a_{02}$\\
$\Phi_1^3$ & $\bar{F}_{\alpha{\beta}\gamma}$ & $a_{03}$\\
$\Phi_2$ & $\bar{H}_{ijk}$ & $a_{10}$\\
$\Phi_3$ & $\bar{Q}_{k}^{\alpha\beta}$ & $a_{20}$\\
$\Phi_2\Phi_1$ & $\bar{H}_{\alpha{jk}}$ & $a_{11}$\\
$\Phi_3\Phi_1$ & $\bar{Q}_{k}^{\alpha{j}}$,$\bar{Q}_{k}^{i\beta}$, $\bar{Q}_{\alpha}^{\beta\gamma}$ & $a_{21}^1$,$a_{21}^2$,$\tilde{a}_{21}^3$\\
$\Phi_2\Phi_1^2$ & $\bar{H}_{i\beta\gamma}$ & $a_{12}$\\
$\Phi_3\Phi_1^2$ & $\bar{Q}_{\gamma}^{i\beta}$,$\bar{Q}_{\beta}^{\gamma{i}}$,$\bar{Q}_{k}^{ij}$ & $a_{22}^1$,$a_{22}^2$,$a_{22}^3$\\
$\Phi_2\Phi_1^3$ & $\bar{H}_{\alpha\beta\gamma}$ & $a_{13}$\\
$\Phi_3\Phi_1^3$ & $\bar{Q}_{\gamma}^{ij}$ & $a_{23}$\\
\hline
\end{tabular}
\end{table}
\end{center}

\section{Algorithm Code}
\label{ap:code}

\begin{mylisting}
\begin{verbatim}
clc
clear all
global a0 a1 a2 a3 b0 b1 b2 b3 c0 c12 c13 c22 c23 c3;
max=1000;                           % Maximum number of vacua in the landscape
Nu=10;                              % Cutoff for the fluxes
nvar=6;                             % Number of variables in the scalar potential
time_ga_parallel=0;
sol=zeros(max,7);
fluxes=zeros(max,14);
fprintf('\n Minimization of moduli begins\n\n');
for i=1:max;
    clc;
    numero=randi(10);
    [a0 a1 a2 a3 b0 b1 b2 b3 c0 c12 c13 c22 c23 c3]=solucion(numero,Nu);
        
    lb=[0,0,0,0,0,0];
     
    V1=@(x) \% Function to minimize
    
    options = gaoptimset('Generations',4000,'TolFun',1e-20,'PopulationSize',100,'TolCon',1e-4,...
    'MutationFcn',@mutationadaptfeasible,'StallGenLimit',35,'Display','off','InitialPopulation',initial,'HybridFcn',@fmincon);
    
    startTime = tic;
    [x,fval,flag]=ga(V1,nvar,[],[],[],[],lb,[],[],options);
    time_ga_parallel = toc(startTime);
    
    sol(i,:)=[x,fval];
    fluxes(i,:)=[a0 a1 a2 a3 b0 b1 b2 b3 c0 c12 c13 c22 c23 c3];
    fprintf('Iteration \%g takes \%g seconds.( \%g, \%g, \%g , \%g , \%g , \%g , \%g) \n',i,time_ga_parallel,sol(i,1),sol(i,2),...
	    sol(i,3),sol(i,4),sol(i,5),sol(i,6),sol(i,7));
end
figure(1);
loglog(1./sol(:,4),-sol(:,7),' x ');
xlabel('g_s');
ylabel('ln(\Lambda )');
\end{verbatim}
\end{mylisting}


\section{A window for inflationary conditions}
\label{ap:mass}
The existence of stable De Sitter vacua, as we have seen in the body of this manuscript, depends on the way SUSY is broken and the specific flux configuration we are considering.  However, from the point of view of cosmology, it is interesting to look for extra conditions, namely: 1) Small values for the vacuum energy (positive or negative) and 2) conditions for the presence of inflation (slow-roll parameters for the effective scalar potential). \\

Here we want to show that there are generic issues which allow us to expect that there are big chances to find  specific models which allows the presence of inflation by assuming small values for vacuum energy (although this requirement is not necessary).\\

Our strategy is to consider a subset of solutions on which the possibility to have an  inflation driven by a subset of the moduli space is much bigger than its absence. This subset of solutions represents the "window" we are going to focus on. The idea is the following: we shall assume that it is possible to identify a subset of the moduli fields as the inflaton. This would require that the masses of such fields are larger than the rest.  If we can show that this is the most probable scenario we will conclude that the identification of  those fields with the inflaton is also a very likely situation. To do that, we will proceed the other way around, i.e., we will show that assuming all masses to be of the same order is a very restrictive option,  while having the contrary represents the big subset of solutions. This argument does not imply the existence of inflation conditions, but open up the possibility to find them.

\subsection{Mass hierarchies}

We shall show that there exists conditions for inflation by probing that  the absence of inflation is very restrictive. Our statement is as follows: If all masses computed from the scalar potential are of the same order, there would not be  a hierarchy on the masses which otherwise would lead to the presence of a prefer inflationary direction driven by a linear combination of moduli fields.\\

Therefore, if constraining the scalar potential terms such that all masses are of the same order in Planck units is not consistent, it would mean that all moduli fields would decay at the same rate and there would not be an inflationary direction on the potential, although a multi-field inflation driven by all moduli could be favored.\\

Let us therefore assume that all masses computed from the scalar potential are of the same order. This statement restricts the terms of the polynomials $P_j(\Phi_1)$. Since the square magnitude of these polynomials appear in the mass terms, their components must be of the same order since all of them depend only on $\Phi_1$. \\

We shall concentrate on the diagonal terms in mass matrix (since this is enough for our purposes).  Let us consider first the masses related to the moduli fields $\Phi_2$ and $\Phi_3$.  It is easy to check that 
\begin{equation}
M^2_{\phi_i\phi_i}=\frac{\partial^2 \mathscr{V}}{\partial \phi_i^2}\sim 2\left(\frac{\psi^2_1}{|\Phi_1|^2}+1\right)|P_i(\Phi_1)|^2,
\end{equation}
for $i=2,3$. If all terms in the polynomial contribute significantly to the mass, this means that
\begin{equation}
{\cal O}(<a_{j0}>)\sim {\cal O}(<a_{j1}\Phi_1>)\sim {\cal O}(<a_{j2}\Phi_1^2>)\sim {\cal O}(<a_{j3}\Phi_1^3>),
\end{equation}
where $<\Phi_1>$ is the value of $\Phi_1$ at the minimum of the potential (if exists).  Consider a generic term $a_{jm}\Phi_1^m$. If such term is present it must be of the same order that $a_{j(m+1)}\Phi_1^{m+1}$. This implies that
\begin{equation}
<a_{jm}>\sim <a_{j(m+1)}\Phi_1>.
\end{equation}
Let us say that $<\Phi_1>\sim 10^{T}M_p$, for a real $T<1$ (since we are assuming small values on the scalar potential). Let us also assume that the real coefficients $a_{j(m+1)}$, related to the integrated fluxes, are all of order $10^K$ in mass Planck units. For the above two terms of the polynomial to be of the same order we find that
\begin{equation}
{\cal O}(<a_{jm}>)\sim 10^{K+T}, 
\end{equation}
in Plank units. Therefore we would require a fractional amount of fluxes violating Dirac quantization. Hence,  $P_j(\Phi_1)$ is restricted to be of the form
\begin{equation}
P_i(\Phi_1)=a_{im}\Phi_1^m,
\label{poly2}
\end{equation}
for some $m\in\{0,1,2,3\}$. Notice that at least one of the polynomials must depend on $\Phi_1$, i.e., for some $i$, $m$ must be different than 0. It follows then that both polynomials $P_2$ and $P_3$ must be of the same order at the minimum of the potential. From expression (\ref{poly}), we observe that this happens if $m_2=m_3$.\\

Now, consider the mass term $M_{\phi_1\phi_1}$, which assuming that ${\cal O}(P_2)\sim {\cal O}(P_3)$ is given by
\begin{equation}
M^2_{\phi_1\phi_1}\sim \big(2\frac{\phi_1\psi_1^2}{|\Phi_1|^4}+1\big)|\mathscr{W}|^2+2m_2~\frac{\phi_1}{|\Phi_1|^2}\big(\psi^2_2|P_2|^2+\psi_3^2|P_3|^2+\psi_2~Im~(\mathscr{W}\bar{P_2})+\psi_3~Im~(\mathscr{W}\bar{P_3})\big).
\end{equation}
This mass term would be of the same order as $M_{\phi_2,\phi_2}$ and $M_{\phi_3,\phi_3}$ if ${\cal O}(\psi_2 P_2)\sim {\cal O}(\psi_3 P_3)\sim {\cal O}(P_1)$. One possibility consists in taking a vanishing $P_1$, but this would imply the absence of NS-NS fluxes violating the tadpole condition. Therefore $P_1$ must be different from zero, leading to the constraint that
\begin{equation}
{\cal O}(<\Phi_1^{m_1}>)\sim {\cal O}(<\psi_2\Phi_1^{m_2}>).
\end{equation}
Assuming that the order of $<\psi_2>$ is $<\Phi_1>^s$ with an arbitrary $s$, we are forced to say that $m_1=s+m_2$. However, since the numbers $m_1$ are the degrees on the polynomial coefficients, they must be  integer numbers.\\

 On the other hand,   we also see from the tadpole condition  that $m_1-m_2=1,3$. Therefore, a scenario where all masses are of the same order yields to a vacuum expectation value for the real part of the dilaton and the K\"alher structure to be of the same order or 3 order of magnitude bigger than the vev of the complex structure. In summary, we find that in order to have all masses of the same order we must fulfill the following conditions:
 
 \begin{enumerate}
 \item
 Small values for $<\Phi_1>$, which lead to small values for the scalar potential (since it is constructed by positive defined terms).
 \item
 All fluxes of the same order. 
 \item
$ {\cal O}(<\Phi_2>)\sim {\cal O}(<\Phi_3>)\sim \left({\cal O}(<\Phi_1>)\right)^{1, 1/3}$. 
 \end{enumerate}
 
This scenario in which inflation guided by a subset of all moduli is not present seems to be quite restrictive due to the condition (3). The possibility that inflation is indeed driven by a subset of the moduli is therefore the biggest one although we can relax the first two points. Relaxing point (1) would lead us to big values for the minima of the scalar potential. The second point can nevertheless be relaxed, but no so far from small values for the fluxes, otherwise we could depart from smooth geometries and backreactions must be considered. Notice that the mass terms above considered corresponds to elements of a non-diagonal mass matrix. Once it is diagonalized by computing the proper values, the order of the diagonal terms keep close to each other.





\newpage
\bibliography{non-geom-cosmology}
\addcontentsline{toc}{section}{Bibliography}
\bibliographystyle{TitleAndArxiv}

\end{document}